\documentclass[fleqn,usenatbib]{mnras}

\usepackage{newtxtext,newtxmath}

\usepackage[T1]{fontenc}

\DeclareRobustCommand{\VAN}[3]{#2}
\let\VANthebibliography\thebibliography
\def\thebibliography{\DeclareRobustCommand{\VAN}[3]{##3}\VANthebibliography}

\usepackage{graphicx}	
\usepackage{amsmath}	
\usepackage{textcomp}	
\usepackage[dvipsnames]{xcolor}
\usepackage{tikz}
\usetikzlibrary{arrows.meta,
				 shapes,
                chains,
                positioning,
                shapes.geometric}
\tikzstyle{arrow}=[draw, -latex]

\newcommand{\appropto}{\mathrel{\vcenter{
  \offinterlineskip\halign{\hfil$##$\cr
    \propto\cr\noalign{\kern2pt}\sim\cr\noalign{\kern-2pt}}}}}

\newcommand{\Ssh}{S_{\zeta,\mathrm{sh}}}
\newcommand{\cs}{c_\mathrm{s}}
\newcommand{\lsh}{l_\mathrm{sh}}
\newcommand{\Mp}{M_\mathrm{p}}
\newcommand{\Mth}{M_\mathrm{th}}
\newcommand{\hp}{h_\mathrm{p}}
\newcommand{\Hp}{H_\mathrm{p}}
\newcommand{\Rp}{R_\mathrm{p}}
\newcommand{\Pp}{P_\mathrm{p}}
\newcommand{\de}{\mathrm{d}}

\newcommand{\pd}[2]{\frac{\partial #1}{\partial #2}}
\newcommand{\td}[2]{\frac{\de #1}{\de #2}}

\newcommand{\azav}[1]{\left\langle #1 \right\rangle_\phi}

\newcommand{\vect}[1]{\mathbf{#1}}
\newcommand{\abs}[1]{\left\vert{#1}\right\vert}

\newcommand{\I}{\mathrm{i}}

\newcommand{\RE}{\mathrm{Re}}
\newcommand{\IM}{\mathrm{Im}}
\newcommand{\F}{\mathcal{F}}
\newcommand{\Am}{\mathcal{A}_m}

\newcommand{\tlr}{t_\mathrm{lin}}
\newcommand{\tlrsa}{t^\mathrm{SA}_\mathrm{lin}}
\newcommand{\tlrsim}{t^\mathrm{sim}_\mathrm{lin}}
\newcommand{\tn}{t_\mathrm{nl}}
\newcommand{\tlsa}{t^\mathrm{SA}_\mathrm{nl}}
\newcommand{\tlsim}{t^\mathrm{sim}_\mathrm{nl}}
\newcommand{\thsim}{\hat{t}^\mathrm{sim}_\mathrm{nl}}

\newcommand\ba{\begin{eqnarray}}
\newcommand\ea{\end{eqnarray}}

\defcitealias{Goodman2001}{GR01}
\defcitealias{Rafikov2002}{Rafikov 2002}
\defcitealias{Cimerman2021}{Paper I}

\title[Emergence of vortices]{Emergence of vortices at the edges of planet-driven gaps in protoplanetary discs}

\author[N. P. Cimerman and R. R. Rafikov]{
Nicolas P. Cimerman$^{1}$\thanks{E-mail: npcphys@gmail.com (NPC)},
Roman R. Rafikov$^{1,2}$
\\
$^{1}$Department of Applied Mathematics and Theoretical Physics, University of Cambridge, Wilberforce Road, Cambridge CB3 0WA, UK\\
$^{2}$Institute for Advanced Study, Einstein Drive, Princeton, NJ 08540, USA
}

\date{Accepted XXX. Received YYY; in original form ZZZ}

\pubyear{2020}

\begin{document}
\label{firstpage}
\pagerange{\pageref{firstpage}--\pageref{lastpage}}
\maketitle

\begin{abstract}
Young planets embedded in protoplanetary discs (PPDs) excite spiral density waves, which propagate, shock and deposit angular momentum in the disc. This results in gap opening around the planetary orbit, even for low (sub-thermal) mass planets, provided that the effective viscosity in the disc is low. The edges of these planet-induced gaps are known to be prone to emergence of observable vortices via the Rossby Wave Instability (RWI). We study timescales for the development of vortices driven by low mass planets in inviscid discs. We employ a recently developed semi-analytical theory of vortensity production by the planet-driven shock to predict vortensity evolution near the planet, from which we derive the radial profile of the planet-induced gap as a function of time (this procedure can have multiple other uses, e.g. to study dust trapping, suppression of pebble accretion, etc.). We then analyze the linear stability of the gap edges against the RWI, obtaining the timescales for the first appearance of unstable modes and (later) fully developed vortices at gap edges. We present useful formulae for these timescales as functions of planetary and disc parameters and provide their physical justification. We also thoroughly test our semi-analytical framework against high resolution 2D hydrodynamic simulations, confirming the accuracy of our theoretical predictions. We discuss ways in which our semi-analytical framework can be extended to incorporate additional physics, e.g. planetary accretion, migration, and non-zero disc viscosity. Our results can be used to interpret observations of PPDs and to predict emergence of vortices in simulations.
\end{abstract}





\section{Introduction}
\label{sec:intro}


Recent observations of protoplanetary discs (hereafter PPDs) have shown a plethora of substructures in the millimetre continuum emission that probes the spacial distribution of large dust grains \citep[e.g][]{Andrews2020}. While many substructures are axisymmetric, ring- or gap-like, there are some that show intriguing non-axisymmetric lobes or arcs \citep{vdM2016,Kraus2017,Dong2018,Perez2018}, which have been interpreted as dust traps inside vortices \citep[e.g.][]{vdM2013,Baruteau2019}. Supporting this interpretation, velocity measurements using CO emission lines have revealed kinematic structures in HD 142527, which have been tentatively attributed to vortices \citep{Boehler2021}. Providing further information, vortices can also reveal themselves in near-infrared scattered light images of discs \citep{Zhu2015,Marr2022}. Beyond these promising observable features, vortices have also been suggested to provide efficient particle traps that may gravitationally collapse and enhance planetesimal formation rates \citep{Meheut2012,Zhu2014,Zhu2014II}. They can also affect orbital migration of planets via their mutual gravitational coupling \citep[e.g][]{Li2009,Yu2010,Lin2010,McNally2019}. These observations and theoretical ideas provide ample motivation to study the origin of vortices in PPDs, especially since they can act as signposts of ongoing planet formation.

Indeed, young planets that are still embedded in their natal PPDs are known to gravitationally excite spiral density waves that carry energy and angular momentum across the disc. Wave angular momentum can be transferred to the background disc material via the linear or non-linear wave damping, driving disc evolution and resulting in gap formation around the planetary orbit \citep{Goodman2001,Rafikov2002,Rafikov2016,Miranda2020I,Miranda2020II}. The edges of deep planet-driven gaps can become susceptible to the Rossby Wave Instability \citep[RWI,][]{Lovelace1999,Li2000}, a linear shear instability associated with growing non-axisymmetric perturbations, that eventually breaks the gap edge into vortices \citep{dvB2007,Li2009,Yu2010,Lin2010}. Thus, vortices can potentially reveal the planets that induce them in the first place. 

The details of the vortex production process depend on the disc and planet properties, as well as the wave damping mechanism. The efficiency of linear wave damping mechanisms depends on disc thermodynamics and effective viscosity \citep{Goodman2001,Miranda2020I,Miranda2020II}, while non-linear damping --- wave steepening due to its non-linearity (a finite amplitude effect) resulting in the formation of a shock --- is robust and unavoidable. Partly for that reason, in this work we will focus on the non-linear wave damping. 

This type of damping is very efficient even for weakly non-linear waves launched by relatively low-mass planets \citep{Goodman2001,Rafikov2002}, with masses $M_\mathrm{p}$ below the so-called \textit{thermal mass}:
\begin{align}
	\Mth=\frac{\cs^3}{\Omega_\mathrm{p} G}=\left(\frac{\Hp}{\Rp}\right)^3 M_\star=\hp^3 \, M_\star,
	\label{eq:Mth}
\end{align}
where $\cs$ is the sound speed, $\Omega_\mathrm{p}$ is the orbital angular frequency at the planetary distance $\Rp$, $\Hp$ and $\hp=\Hp/\Rp$ are the disc scale height and aspect ratio at $\Rp$, and $M_\star$ is the stellar mass. It has been shown that even such {\it sub-thermal} planets can be responsible for producing prominent features in protoplanetary discs, both gaps/rings  \citep{Dong2017,Bae2017,Zhang2018, Miranda2019II,Miranda2020I,Miranda2020II} and vortices \citep{Hallam2020,Hammer2021}. For this reason, as well as to enable (semi-)analytical progress in understanding vortex formation, in this work we will focus on planets with $M_\mathrm{p}\lesssim \Mth$.

Non-linear damping of planet-driven density waves proceeds via their inevitable evolution into shocks. Planetary shock fronts are known \citep{Kevlahan1997,Lin2010,Dong2011II} to produce a jump of {\it vortensity} (or potential vorticity) $\zeta=\Sigma^{-1}\mathbf{\nabla}\times \mathbf{u}$ (here $\Sigma$ is the disc surface density and $\mathbf{u}$ is the fluid velocity), a quantity that is otherwise conserved along streamlines in two-dimensional (2D), inviscid and barotropic flows. As fluid elements periodically cross the planet-induced shock, disc vortensity near the planet steadily evolves, a process closely related to gap opening \citep{Muto2010}. Eventually, this vortensity evolution triggers the RWI \citep{Lovelace1999} and vortices emerge in the disc.

The timescale on which this pathway for vortex formation operates is an important diagnostic, which can help us constrain the properties of putative planets triggering vortices. Indeed, higher mass planets launch stronger density waves, which drive faster vortensity evolution and cause vortices to form earlier. Thus, if a PPD is observed to have a gap suggestive of a planet opening it and features a vortex at the gap edge, this would imply that either (i) planetary mass is high or (ii) the planet has been present for a long time. Both constraints are useful for interpreting observations. However, so far the problem of determining the vortex emergence timescale has received little attention\footnote{\citet{Hallam2020} studied the relation between the planetary growth timescale and the emergence of vortices, but they also varied disc viscosity and adopted a particular accretion history of the planet.}, with most studies focusing on the long-term survival of already well-developed vortices under various secondary instabilities \citep{Lesur2009}, thermal relaxation \citep{FungOno2021,Rometsch2021} and other processes.

The main goal of our present study is to determine the critical time for the appearance of planet-driven vortices in inviscid discs and to explore the dependence of this timescale on the planetary and disc parameters, which are fixed in time. We do this using both direct hydrodynamic simulations and semi-analytical theory based on the recent work of  \citet[][hereafter \citetalias{Cimerman2021}]{Cimerman2021}, in which we developed from first principles (and verified numerically) a semi-analytical framework for calculating the vortensity jump $\Delta \zeta$ across a planet-driven spiral shock in a barotropic disc, taking disc and planet parameters as inputs. This framework  as well as the linear stability analysis of the planet-induced gaps against the RWI are the key theoretical foundations of our present study. We also perform a thorough test of our method against direct, high resolution, 2D, hydrodynamic simulations for the range of relevant parameters, allowing us to verify its validity and identify its limitations.

This paper is organised as follows (readers who are only interested in results for the timescales and their applicability might skip to Section \ref{sec:timescales}). We describe the problem setup in Section \ref{sec:setup}, followed by a description of non-linear simulations and their typical outcomes in Section
\ref{sec:hydro-sim}. In Section \ref{sec:RWIlin} we introduce the linear stability analysis for RWI at gap edges, explain our method for obtaining the disc profile from its vortensity structure and show typical results for unstable modes. Relevant timescales for different stages of developing RWI are introduced and motivated, together with related diagnostics in Section \ref{sec:diag}. In Section \ref{sec:timescales} we show results for these timescales and provide power law fits, propose a heuristic theoretical explanation for these dependencies and study the vortensity levels at the onset of RWI. We discuss our findings, their applications and limitations in Section \ref{sec:discuss} before  summarizing in Section \ref{sec:summ}. Appendices contain further information and tests regarding our methods.


\section{Problem setup}
\label{sec:setup}


Our setup for studying planet-disc interaction is the same as in \citetalias{Cimerman2021}. Here, we recite its most important aspects.

We consider a planet of mass $M_\mathrm{p}$ orbiting a star of mass $M_\star$ on a circular orbit with a semi-major axis $R_\mathrm{p}$ that lies within a thin two-dimensional gas disc. We adopt polar coordinates $(R,\phi)$. The initial (background) disc state, unperturbed by the planet has a power law profile of the surface density
\begin{align}
	\Sigma_\mathrm{i}(R) = \Sigma_\mathrm{i}(R_\mathrm{p}) \left(\frac{R}{R_\mathrm{p}}\right)^{-p} = \Sigma_\mathrm{p} \left(\frac{R}{R_\mathrm{p}}\right)^{-p},
\end{align}
where $p$ is a constant.\footnote{We have changed the subscript for the initial conditions from '0' to 'i' as compared to \citetalias{Cimerman2021}, to avoid confusion with azimuthally averaged variables, i.e. their $m=0$ Fourier components.} We assume that the mass of the disc is small, $M_\mathrm{d} \ll M_\star$, such that its self-gravity can be neglected. 

We use a {\it globally isothermal} equation of state (EoS), $P = \cs^2 \Sigma$, with the spacially constant speed of sound $\cs$, which is barotropic (i.e. $P = P(\Sigma)$ only). The choice of this EoS instead of an often used non-barotropic \textit{locally} isothermal EoS, for which the sound speed follows a prescribed radial profile $\cs(R)$, is motivated in more detail in Section \ref{subsec:therm} and \citetalias{Cimerman2021}. The 2D approximation is appropriate for thin discs, such that the aspect-ratio of the disc $h = H/R = \cs/(\Omega R) \ll 1$.

Very importantly, in this study \citep[and unlike e.g.][]{Hammer2017,McNally2019,Hallam2020} we keep both planetary ($\Mp$ and $\Rp$) and disc ($\hp$) properties fixed in time to highlight the key physical processes and to reduce the number of relevant parameters. In Section \ref{subsec:appl} we comment on how this assumption can be relaxed.

In general, the disc maintains a radial centrifugal balance accounting for the radial pressure gradient:
\begin{align}
	\Omega^2(R) = \Omega_\mathrm{K}^2(R) + \frac{\cs^2}{R \Sigma(R)} \td{\Sigma(R)}{R},
	\label{eq:radHSE}
\end{align}
where $\Omega_\mathrm{K} = \sqrt{GM_\star/R^3}$ is the Keplerian orbital frequency. The initial, unperturbed (by either a planet or vortices) disc has radial velocity $u_{\mathrm{i},R}(R) = 0$ and azimuthal velocity $u_{\mathrm{i},\phi}(R) = R\Omega_\mathrm{i}$, where $\Omega_\mathrm{i}(R)$ is given by equation (\ref{eq:radHSE}) with $\Sigma(R)=\Sigma_\mathrm{i}(R)$.
 
Our fiducial disc model is the same as in \citetalias{Cimerman2021}. It has an aspect-ratio $h_\mathrm{p} = 0.05$ at $\Rp$ and a surface density slope $p = 3/2$. The latter results in an initial vortensity profile $\zeta_\mathrm{i}(R)$ that is almost constant for slightly sub-Keplerian discs, see Section 6 of \citetalias{Cimerman2021}. 

When presenting our results we adopt units where $G = M_\ast = \Omega_K(R_\mathrm{p}) = \Sigma_\mathrm{p} = 1$. In these units the planetary orbital time is $P_\mathrm{p} = 2\pi$, which we use as a unit of time.

We now describe the different methods, both numerical and analytical, used in this work (their usage is illustrated in Fig. \ref{fig:flowchart}). We use the fiducial setup described earlier to showcase our methods and typical results.


\section{Methods: hydrodynamical simulations}
\label{sec:hydro-sim}


\begin{figure*}
\centering		
\includegraphics[width=0.99\textwidth]{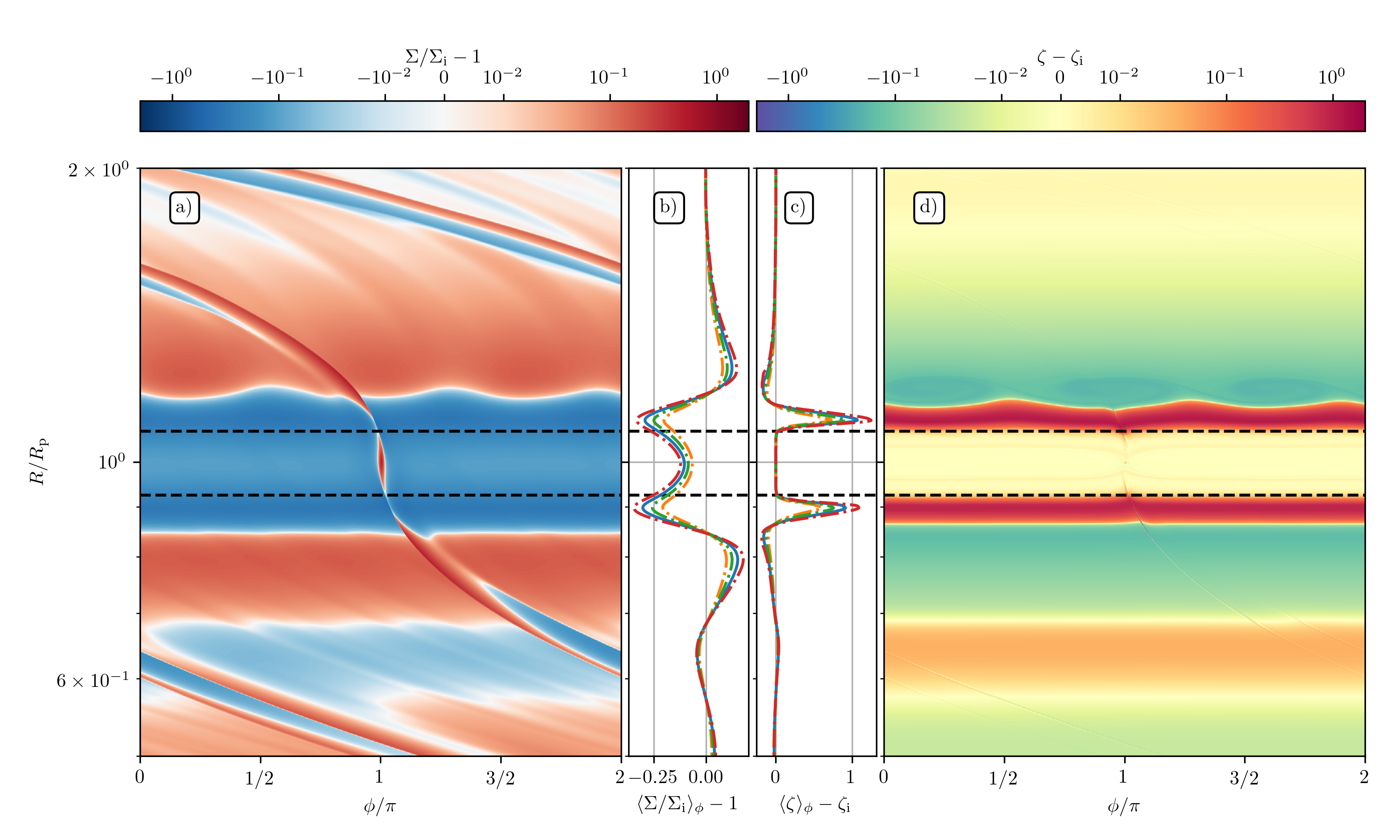}
\vspace*{-0.1cm}
\caption{Simulation results using fiducial disc parameters and an intermediate-mass planet $M_\mathrm{p}/M_\mathrm{th} = 0.25$. All perturbations are w.r.t. the initial conditions ($\Sigma_\mathrm{i}, \zeta_\mathrm{i}$). (a) Map of the relative surface density perturbation. (b) Radial profile of the azimuthally averaged surface density perturbation. (c) Radial profile of the azimuthally averaged vortensity perturbation. (d) Map of the vortensity perturbation. The dashed horizontal lines show the predicted shock locations $R = R_\mathrm{p} \pm l_\mathrm{sh}$. At the outer gap edge, the RWI has developed and led to the formation of three large vortices. These cause additional, weaker spiral waves that are visible in panel (a). 2D maps and blue solid lines in panels (b) and (c) show the disc state at $t = 780 \Pp$. Additional dot-dashed lines in the middle panels
correspond to $t = \lbrace 500, 640, 920\rbrace \Pp$.
}
\label{fig:dvort_map}
\end{figure*}

\begin{figure*}
\centering		
\includegraphics[width=0.99\textwidth]{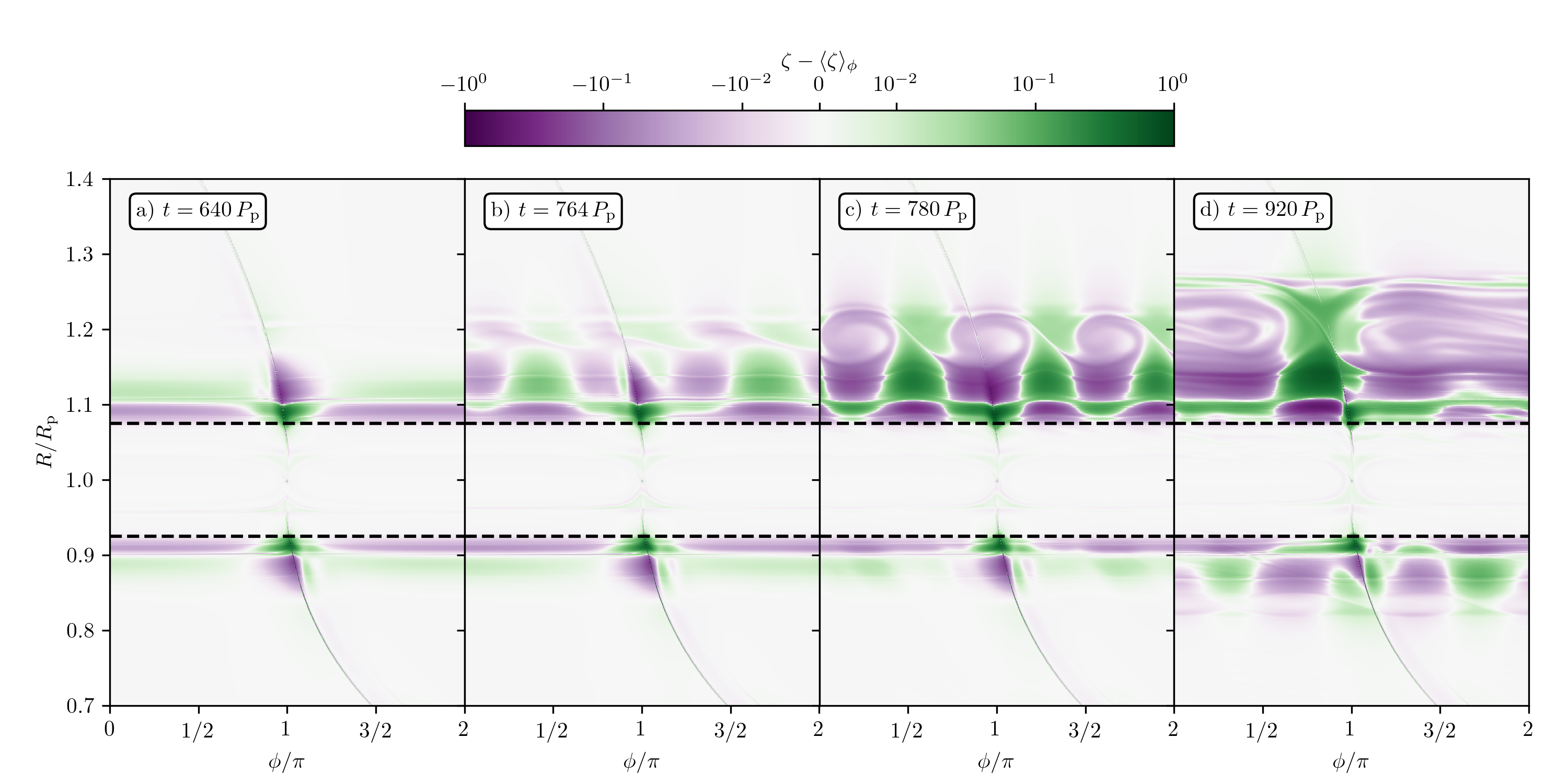}
\vspace*{-0.1cm}
\caption{
Two-dimensional map of the non-axisymmetric part of the vortensity perturbation, $\zeta - \azav{\zeta}$, for the fiducial case. Panels show results for different times (labels in panels) during the onset of instability. The shock front is visible as a thin curve.
In panel (a) the flow is stable and mostly axisymmetric. At the predicted distance $\lsh$ from the planet (black dashed lines), the shock starts leading to roughly axisymmetric vortensity distribution --- two thin rings around $R \simeq 0.9$ and $R \simeq 1.1$ (locally perturbed around the shock). Other panels show (b) the onset of RWI in its initial stage, (c) the development of three vortices and (d) a single vortex as their merger product at the outer gap edge. At the inner gap edge, we observe similar processes starting later.
}
\label{fig:dvort_azav_map}
\end{figure*}


\subsection{Setup and overview of simulations}


To study vortex formation we perform global, non-linear hydrodynamic simulations of planet-disc interaction using Athena++ \citep{Athenapp2020}\footnote{Athena++ is publicly available on \href{https://github.com/PrincetonUniversity/athena}{GitHub}.}. The code uses a Godunov scheme to solve the hydrodynamic equations in a conservative form:
\begin{align}
    \pd{\rho}{t} + \nabla \cdot (\rho \vect{u}) &= 0, \\
    \pd{(\rho \vect{u})}{t} + \nabla \cdot (\rho \vect{u} \otimes \vect{u} + P \vect{I}) &= - \rho \nabla \Phi,
\end{align}
where $P$ is the gas pressure, $\vect{I}$ the identity tensor, and $\Phi$ is the total gravitational potential.
Following \citetalias{Cimerman2021}, we employ a fourth-order smoothed potential for the planet $\Phi_\mathrm{p} = \Phi_\mathrm{p}^{(4)}$ \citep{Dong2011I}. 

The simulation domain extends over the radial range $0.2 \leq R/R_\mathrm{p} < 4.0$, with logarithmic grid spacing. The full azimuthal range $0 \leq \phi \leq 2\pi$ is covered uniformly. In this work, we use a fixed resolution of $N_R \times N_\phi = 3448 \times 7200$, corresponding to 58 cells per disc scale-height at $\Rp$ for the fiducial disc with $\hp =0.05$. For hotter discs (larger $\hp$) this results in higher effective resolution. We use Roe's approximate Riemann solver, linear spacial interpolation and second order accurate time-stepping. We employ the orbital advection algorithm, which has recently been implemented in Athena++. For more details regarding the numerical setup, see \citetalias{Cimerman2021}.


\subsection{Typical simulation outcomes}

We now describe typical outcomes of our simulations. Our fiducial model, which we will use as a reference, has $\Mp = 0.25 \Mth$, $\hp = 0.05$ and $p=1.5$ (i.e.  $\zeta_\mathrm{i}$ close to a constant). We illustrate our results with 2D snapshots of the surface density and vortensity perturbations at a particular moment of time, $t = 780 \Pp$, in panels (a) and (d) of Fig. \ref{fig:dvort_map}. In panels (b) and (c) of that figure we also show the azimuthally-averaged profiles of the $\Sigma$ and $\zeta$ perturbations at several moments of time to illustrate their steady evolution. Additionally, we display maps of $\zeta - \azav{\zeta}$, i.e. the non-axisymmetric part of the vortensity, in Fig. \ref{fig:dvort_azav_map}, which makes it easier to identify vortices as they develop. 

For more than 700 orbits of the simulation, the vortensity perturbation, driven by the spiral shocks, increases (close to) linearly with time, maintaining an almost constant $\partial_t \azav{\zeta}(R)$ and showing no sign of saturation  (see also Fig. 12  in \citetalias{Cimerman2021}). As expected, vortensity stays almost unperturbed in the radial band $\vert R-\Rp \vert < \lsh$ (shown with horizontal dashed lines), where \citep{Goodman2001}
\begin{align}
    \lsh\approx 0.86\,\hp\,\Rp\left(\Mp/\Mth\right)^{-2/5}
    \label{eq:lsh}
\end{align}
(for isothermal EoS) is the wave shocking length; after travelling this distance the wave shocks, driving time evolution of $\zeta$ for $\vert R-\Rp \vert > \lsh$. This evolution, visible in Fig. \ref{fig:dvort_map}c, is associated with the formation of a gap that becomes deeper and deeper (see Fig. \ref{fig:dvort_map}b), leading to increasing radial gradients of $\Sigma$ and pressure. This, in turn, leads to increased radial shear of the azimuthal flow, which nevertheless remains stable (laminar) during this stage. This can be seen in Fig. \ref{fig:dvort_azav_map}a, in which the only non-axisymmetric vortensity perturbation is associated with the planetary spiral shock. 

This picture begins to change at around $t = 760 \Pp$, when our simulation starts exhibiting RWI at the outer gap edge ($R/\Rp \simeq 1.1$), which is clear from the non-axisymmetric vortensity structures appearing in Fig. \ref{fig:dvort_azav_map}b.
As the instability sets in, non-axisymmetric features also become apparent in the surface density and vortensity perturbations in Fig. \ref{fig:dvort_map}a,d slightly later, at $t=780 \Pp$; the corresponding $m=3$ vortensity perturbation at the outer gap edge is also clear in Fig. \ref{fig:dvort_azav_map}c, appearing as three vortices. These vortices drive additional spiral density waves, visible in the inner and outer disc, that are weaker than the planet-driven density waves (see also Fig. 1 in \citetalias{Cimerman2021}, when vortices have not formed yet).

At $t=830$ orbits, the multiple vortices have coalesced to form one single $m=1$ vortex. At the inner gap edge ($R/\Rp \simeq 0.9$), the RWI sets in at around $t=880 \Pp$, resulting in merging vortices. This state is illustrated in Fig. \ref{fig:dvort_azav_map}d. By $t = 1000 \Pp$, both gap edges feature a single vortex. The instability smooths out radial gradients of vortensity (and related fluid variables), but does not eliminate their extrema \citep[see also][]{Meheut2010}.


\section{Methods: linear analysis of RWI development}
\label{sec:RWIlin}


In addition to numerical simulations capturing the development of the RWI directly, we also studied the instability using semi-analytical means. Below we present our setup for linear stability analysis of an axisymmetric, 2D barotropic and inviscid flow. 


\subsection{Linear RWI stability analysis}
\label{sec:RWI_eqs}


Consider a barotropic flow in an axisymmetric 2D disc in radial equilibrium with $u_{0,R} = 0$ everywhere, with prescribed radial profiles of surface density and azimuthal velocity, $\Sigma_0(R)$ and $u_{0,\phi}(R) = R \Omega_0(R)$
\footnote{We note that this axisymmetric state is in general different from the initial conditions we use for disc-planet interactions, which are denoted $\Sigma_0$, $\Omega_0$ in 
\citetalias{Cimerman2021} and $\Sigma_\mathrm{i}$, $\Omega_\mathrm{i}$ in this work.}.

Introducing small, non-axisymmetric perturbations to ${X \in [\Sigma,u_R,u_\phi]}$ of the form
\footnote{The perturbations $\delta X = X - X_0$ are perturbations to an axisymmetric state taken at some time $t$ after introduction of the planet and are different from the perturbations with respect to the initial state (i.e. $X - X_\mathrm{i})$.}
\begin{align}
	\delta  X(R,\phi,\omega,m,t) = \delta  X_m(R) \exp \left[\I (m\phi - \omega t) \right],
	\label{eq:RWI_pert}
\end{align}
($\abs{\delta  X} \ll \abs{X_0}$) where $\delta  X_m(R)$ is a complex amplitude, $m$ is the azimuthal mode number and 
\begin{align}
	\omega = \omega_\mathrm{R} + \I \gamma
\end{align}
is the complex mode frequency, with real part $\omega_\mathrm{R}$ and imaginary part $\gamma$ (\textit{unstable} modes have $\gamma > 0$ and grow exponentially with time).

As shown in \citet{Lovelace1999}, the linearised perturbation equations (ignoring the planetary potential) can be combined into a master equation
\begin{align}
	\Psi_m^{\prime \prime}+B(R, \omega, m) \Psi_m^{\prime}+C(R, \omega, m) \Psi_m=0,
	\label{eq:RWI}
\end{align}
where primes stand for radial derivatives and
\begin{align}
	\Psi_m \equiv \frac{\delta  P_m}{\Sigma_0},
\end{align}
is the enthalpy perturbation ($\Psi_m$ is its Fourier component), equal to $\cs^2 \delta  \Sigma_m /\Sigma_0$ in the barotropic case considered here. The coefficients in equation (\ref{eq:RWI}) are related to the disc structure and are given by
\begin{align}
B(R, \omega, m) &\equiv \frac{1}{R}+\frac{\F^{\prime}}{\F}-\frac{\Omega_0^{\prime}}{\Omega_0}, 
\label{eq:B}\\
C(R, \omega, m) &\equiv - \frac{m^2}{R^2} -\frac{\kappa_0^{2}-\Delta \omega^{2}}{c_0^{2}}-2 \frac{m}{R} \frac{\Omega_0}{\Delta \omega} \frac{\F^{\prime}}{\F},
\label{eq:C}\\
\F(R, \omega, m) &\equiv \frac{\Sigma_0 \Omega_0}{\kappa_0^{2}-\Delta \omega^{2}}.
\label{eq:F}
\end{align}
Here $\Delta \omega \equiv \omega -m \Omega_0$ is the Doppler-shifted mode frequency and
\begin{align}
	\kappa_0^2 = \frac{2 \Omega_0}{R} \td{(R^2 \Omega_0)}{R},
	\label{eq:kappa}
\end{align}
is the square of the epicyclic frequency. Note that in an axisymmetric 2D disc in radial force equilibrium (with $u_R = 0$ everywhere) vortensity can be written as
\begin{align}
	\zeta_0(R) = \frac{1}{R \Sigma_0} \td{(R^2 \Omega_0)}{R}=\frac{\kappa^2_0}{2 \Sigma_0 \Omega_0}.
	\label{eq:vortensity}
\end{align}
Our strategy for numerically solving equation (\ref{eq:RWI}) is described in Appendix \ref{app:find}.

Extremal values in the radial vortensity profile are a necessary condition for the development of the RWI in barotropic discs \citep{Lovelace1999}, and \citet{Pap1989} argued that it is the vortensity {\it minima} that become unstable. The edges of planet-induced gaps in protoplanetary discs can naturally provide such extrema that may become linearly unstable \citep[e.g.][]{dvB2007,Li2009}. We can expose the background vortensity $\zeta_0 = \kappa_0^2/(2 \Sigma_0 \Omega_0)$ in the coefficients (\ref{eq:B})-(\ref{eq:F}) by noting that
\begin{align}
	\F^{-1} &= \frac{\kappa_0^2 - \Delta\omega^2}{\Sigma_0 \Omega_0} = 2\zeta_0 - \frac{\Delta\omega^2}{\Sigma_0 \Omega_0}.
\end{align} 
For corotating modes with $\vert \Delta \omega \vert^2 \ll \vert\kappa_0\vert^2$, one has
$\F^{-1} \simeq 2 \zeta_0$ \citep{Lovelace1999}.

Defining a new function
\begin{align}
	\Xi(R) \equiv \sqrt{\frac{R \F}{\Omega_0}} \Psi,
\end{align}
one can also transform equation (\ref{eq:RWI}) into the form
\begin{align}
	\Xi_m^{\prime \prime} - D(R,\omega,m) \Xi_m = 0
	\label{eq:mastereq}
\end{align}
resembling the time-independent Schr\"odinger equation with complex $\Xi_m$ and 
\begin{align}
 D (R,\omega,m) \equiv \frac{B^{\prime}}{2}+\frac{B^{2}}{4}-C
\label{eq:D}
\end{align}
playing the role of the potential \citep{Ono2016}. Regions of $\RE(D) < 0$ are classically allowed for wave propagation, while regions of $\RE(D) > 0$ are classically forbidden regions, through which waves can only tunnel. Sufficiently deep and wide troughs in $\RE(D)$ allow for trapped modes.

Due to our barotropic equation of state, the Brunt-V\"ais\"al\"a frequency associated with buoyancy vanishes ($N^2 = 0$),
such that stability against axisymmetric perturbations is dictated by the Rayleigh criterion \citep[e.g.][]{Chandra1961}, i.e. $\kappa_0^2< 0$ is needed for instability. On the other hand, equation (\ref{eq:vortensity}) makes it clear that $\kappa_0^2 < 0$ requires $\zeta_0 < 0$. We have checked that all disc models studied in this work remain stable against axisymmetric perturbations (i.e. the minima of vortensity never become negative), according to this criterion. In agreement with previous works, we find that the RWI always occurs before the Rayleigh criterion is violated \citep{Li2000,Les2015}.

Equations (\ref{eq:RWI})-(\ref{eq:D}) make it clear that the linear analysis of RWI requires the knowledge of radial profiles of $\Sigma_0$, $\Omega_0$. These could be derived from simulations, but for a fully self-contained semi-analytical analysis one would like to obtain $\Sigma_0(R)$, $\Omega_0(R)$ without simulations. Next we propose an alternative semi-analytical procedure for constructing $\Sigma_0(R)$, $\Omega_0(R)$ profiles. 

For the rest of this work, we will drop the subscript on $\Sigma_0(R)$, $\Omega_0(R)$ for brevity, with implicit understanding that  $\Sigma(R)$, $\Omega(R)$, etc. refer to the current azimuthally-averaged disc characteristics.


\subsection{Reconstruction of surface density and rotation profiles}
\label{sec:SA_disc}

In \citetalias{Cimerman2021} we developed a semi-analytical procedure for predicting the planetary shock-driven evolution of the disc vortensity $\zeta(R,t)$ in time and space. Now we describe how to use $\zeta(R,t)$ obtained using this procedure and the method outlined in \citet{Lin2010} to reconstruct the surface density and rotation profiles in the disc as a function of time.


\subsubsection{Theoretical considerations}
\label{sssec:comp_sa_sim}

Knowing the vortensity jump at the planet-driven shock $\Delta \zeta (R)$ \citepalias{Cimerman2021}, we can calculate the associated rate of change of the disc vortensity $\Ssh$ according to the following formula:
\begin{align}
	\left.\pd{\zeta(R,t)}{t}\right\vert_\mathrm{sh} \equiv \Ssh = \Delta \zeta (R) \times \frac{\vert \Omega(R) - \Omega_\mathrm{p} \vert}{2 \pi}.
	\label{eq:syn}
\end{align}
Assuming $\Omega(R) \simeq \Omega_\mathrm{K}$, approximating $\Ssh$ as constant in time \citepalias[see][regarding the validity of this assumption]{Cimerman2021}, and integrating over time we obtain the vortensity profile at
time $t$ after the introduction of the planetary potential as
\begin{align}
\zeta(R,t) = \zeta(R,t=0) + t \Ssh(R).
\label{eq:zeta_t}
\end{align}

Substituting $\Omega$ from the relation (\ref{eq:radHSE}) into Equation (\ref{eq:vortensity}) and assuming a barotropic EoS one obtains
\begin{align}
	\zeta = \frac{\Omega_\mathrm{K}^2 + \frac{3 \cs^2}{R} \td{\ln \Sigma}{R} + \cs^2 \td{^2 \ln \Sigma}{R^2}}{2 \Sigma \left( \Omega_\mathrm{K}^2 + \frac{\cs^2}{R} \td{\ln \Sigma}{R}\right)^{1/2}}.
\end{align}
Following \citet{Lin2010}, but using our \textit{globally} isothermal EoS, $\cs =$ const, instead of their \textit{locally} isothermal EoS, we can rearrange this into a second order (non-linear) differential equation for $\Sigma$ similar to their equation (18):
\begin{align}
		\frac{1}{R^3} \td{}{R} \left( R^3 \td{\ln \Sigma}{R} \right) + \frac{\Omega_\mathrm{K}^2}{\cs^2}
		= \frac{2 \Sigma}{\cs^2} \zeta \left( \Omega_\mathrm{K}^2 + \frac{\cs^2}{R} \td{\ln \Sigma}{R} \right)^{1/2},
	\label{eq:ODEsigma}
\end{align}
where vortensity enters only as a source term on the right-hand side. 

This differential equation can be solved numerically in $R$, with the input vortensity profile $\zeta(R,t)$ given by equation (\ref{eq:zeta_t}) at any $t$. To this end, we use an iterative relaxation scheme and impose the unperturbed surface density ($\Sigma = \Sigma_\mathrm{i}$) as a boundary condition far from the planet (see Appendix \ref{app:vortHSE} for more details). 

Having found $\Sigma(R)$, we then retrieve the rotation profile $\Omega(R)$ using equation (\ref{eq:radHSE}), thus fully determining the axisymmetric disc state (i.e. $\zeta(R) \rightarrow [\Sigma(R),\Omega(R)])$. According to \citetalias{Cimerman2021}, our recipe for constructing $\zeta$ is accurate for $0.05 \leq \Mp/\Mth < 1$, implying that the method for constructing time-dependent $\Sigma(R),\Omega(R)$ profiles for discs with gaps should work well for intermediate-mass planets.

In this work we use a fixed planet orbital radius $\Rp$ and planet mass $\Mp$ (resulting in time independent $\Ssh$), but we note that our 1D method can easily include planetary growth and a prescribed planet migration by making these parameters (and $\Ssh$) functions of time (see also Section \ref{subsec:appl}), in which case the second term in (\ref{eq:zeta_t}) will no longer be linear in $t$.


\subsubsection{Verification of $\Sigma(R),\Omega(R)$ retrieval via simulations}
\label{sec:comp_sa_sim}

\begin{figure}
\centering
	\includegraphics[width=0.49\textwidth]{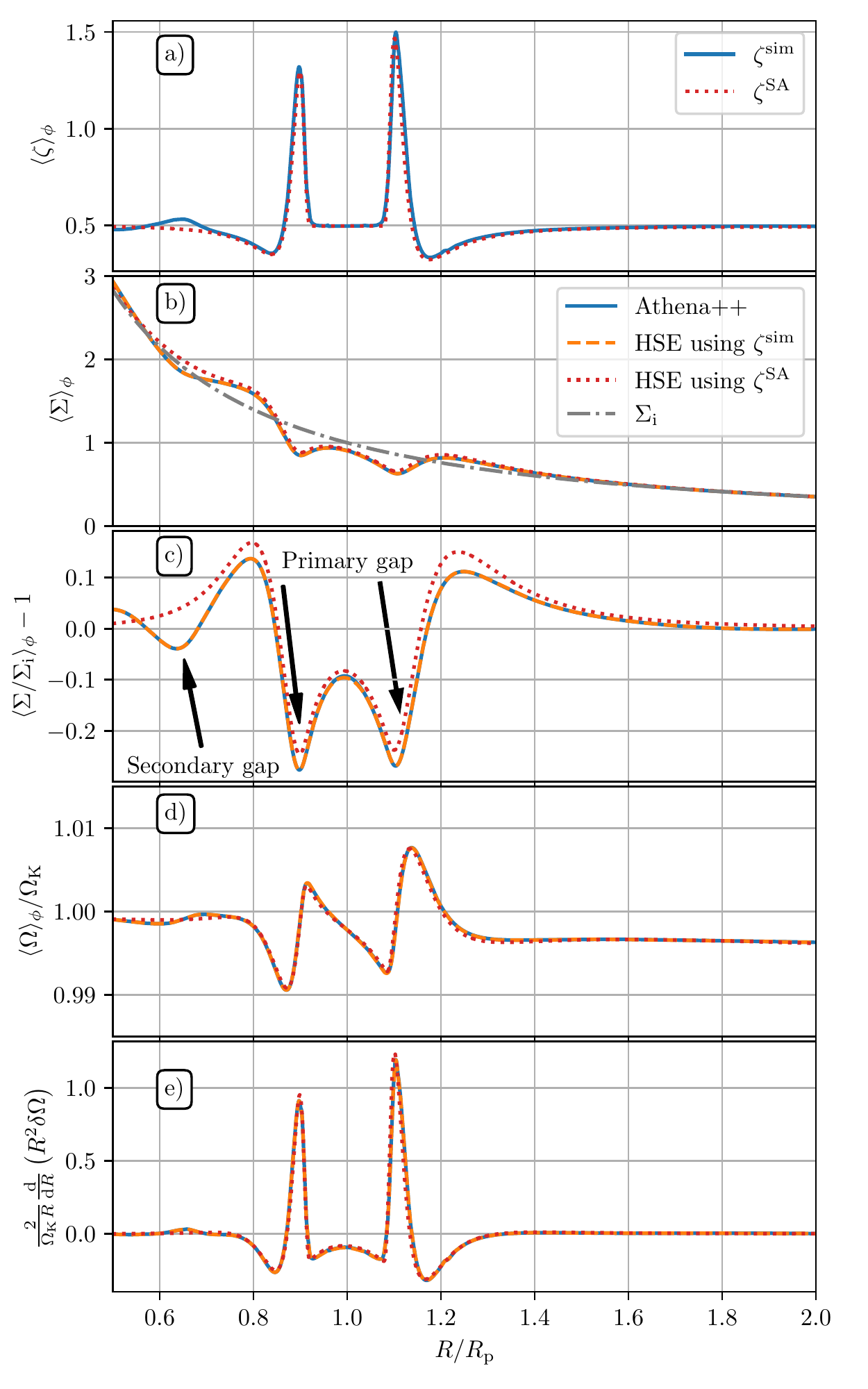}
\vspace*{-0.7cm}
\caption{Radial profiles of (a) vortensity $\langle \zeta \rangle_\phi$ (in units of $\Omega_\mathrm{p}/\Sigma_\mathrm{p}$), (b) surface density (normalized by $\Sigma_\mathrm{p}$) and (c) its relative deviation from $\Sigma_\mathrm{i}$, (d) angular frequency and (e) a measure of shear --- the second term in equation (\ref{eq:zeta_dev}). Shown is the calculation for the fiducial disc model with $\Mp=0.25\Mth$ at $t = 700 \Pp$. We show results from the full hydro simulation (blue) and compare them to results of application of our $\Sigma$, $\Omega$ retrieval procedure (Section \ref{sssec:comp_sa_sim}) with the vortensity profile from simulations (orange dashed) and from our semi-analytical prescription (red dotted). The good agreement between the full simulation results and our semi-analytical procedure is obvious. 
}
\label{fig:rec_disc_q15}
\end{figure}

We now check the performance of the $\Sigma(R),\Omega(R)$ retrieval procedure described in Section \ref{sec:SA_disc} using simulations, by comparing the gap structures obtained by both methods. In Fig. \ref{fig:rec_disc_q15} we show the results for $\Sigma(R),\Omega(R)$ at $t = 700 \Pp$ derived from simulations and through our semi-analytical method for our fiducial disc model. 

Panel (a) shows that the semi-analytical prediction $\zeta^\mathrm{SA}$ (red dotted) provides an excellent match to the vortensity $\zeta^\mathrm{sim}$ (blue solid) obtained with Athena++, reliably reproducing the amplitude and shape of the two vortensity rings. Slight disagreement is present in the inner disc ($R \lesssim 0.7 \Rp$), where $\zeta^\mathrm{SA}$ smoothly tends to the unperturbed value, but the simulation shows a secondary bump and trough. This is caused by the emergence of a secondary spiral arm in the simulation that eventually shocks \citep{Bae2018,Miranda2019I}. Our semi-analytical model cannot capture this effect by design \citepalias[see also][]{Cimerman2021}.

The azimuthally-averaged $\Sigma(R)$ and its relative deviation from the initial $\Sigma_\mathrm{i}(R)$ profile, obtained by the two methods, are shown in panels (b) and (c), respectively. As a test of our disc reconstruction method, we also show the results obtained through the procedure in Section \ref{sec:SA_disc} but with $\zeta^\mathrm{sim}(R,t)$ obtained from simulations (orange dashed), rather than from equation (\ref{eq:zeta_t}). These show excellent agreement with azimuthally averaged simulation results, proving the validity of the $\Sigma(R),\Omega(R)$ retrieval technique (see also \citealt{Lin2010}). Note the characteristic double-trough profile of the gap, predicted in \citet{Rafikov2002II}, with the two minima of $\Sigma$ corresponding to the density wave shocking on each side of the planetary orbit.

When using $\zeta^\mathrm{SA}$ instead of $\zeta^\mathrm{sim}$ to solve equation (\ref{eq:ODEsigma}), the secondary surface density depression and enhancement in the inner disc are missed by $\Sigma^\mathrm{SA}$, corresponding to the missing bump and trough in $\zeta^\mathrm{SA}$. This leads to a smaller gradient at the inner gap edge for the semi-analytical gap. Also, in the entire domain, the semi-analytical solution for $\Sigma(R)$ shows a positive offset with respect to simulation results. Nevertheless, the amplitude and width of the primary bumps are matched reasonably well. In the outer disc, the agreement of the surface density depression and enhancement is better due to the absence of a secondary shock. 

Despite these discrepancies in $\Sigma(R)$, the semi-analytical and numerical radial profiles of the orbital frequency $\Omega(R)$ agree remarkably well, see panel (d). For this parameter set, the outer gap edge shows stronger shear (gradient of $\Omega$) and, quite importantly for the validity of our semi-analytical analysis, $\Omega$ behaviour in the regions of strongest shear at the gap edges is reproduced very well by our method. 

According to the definition of vortensity, its deviation from the initial value $\delta\zeta=\zeta-\zeta_\mathrm{i}$ depends on perturbations of both $\Sigma$ and $\Omega$, i.e. on $\delta\Sigma=\Sigma-\Sigma_\mathrm{i}$ and $\delta\Omega=\Omega-\Omega_\mathrm{i}$. One can easily show that as long as $\Omega_\mathrm{i}\approx\Omega_\mathrm{K}$, we can express
\begin{align}
\frac{\delta\zeta}{\zeta_\mathrm{i}} & \approx\left(1+\frac{\delta\Sigma}{\Sigma_\mathrm{i}}\right)^{-1}\left[1+\frac{2}{\Omega_\mathrm{K}R}\td{}{R}\left(R^2\delta\Omega\right)\right]-1
\label{eq:zeta_dev0}\\     &\approx-\frac{\delta\Sigma}{\Sigma_\mathrm{i}}+\frac{2}{\Omega_\mathrm{K}R}\td{}{R}\left(R^2\delta\Omega\right),      
\label{eq:zeta_dev}
\end{align}
with the second line valid when $\vert\delta\Sigma\vert\ll\Sigma_\mathrm{i}$ and $\vert\partial\delta\Omega/\partial R\vert\ll\Omega_\mathrm{K}/R$; equation (\ref{eq:zeta_dev0}) holds even when these constraints are not met. The behaviour of the second term in the right-hand side of (\ref{eq:zeta_dev}) is illustrated in panel (e), while the first one is shown in panel (c). One can see that even though $|\delta\Omega|/\Omega_\mathrm{K} \lesssim 10^{-2}$ is rather small, see panel (d), the term in panel (e) has a much higher amplitude $\sim 1$; approximation (\ref{eq:zeta_dev}) leads to substantial deviations near the peaks of $\zeta$ in this case. This term considerably exceeds $\delta\Sigma/\Sigma_\mathrm{i}$ in magnitude, which implies that a proper calculation of $\zeta$ evolution must account for the changes in both $\Sigma$ and $\Omega$.

To summarize, our $\Sigma$, $\Omega$ retrieval procedure shows good performance when tested against simulations. In Appendix \ref{sec:tests-SA} we provide further tests of the robustness of our semi-analytical reconstruction for various disc parameters ($p$ and $\hp$) and planetary masses, again finding good agreement with simulations.


\subsection{Illustration of our RWI analysis}
\label{sec:resRWI}


We now show typical results obtained using linear stability analysis on gap profiles produced both by simulations and by the semi-analytical method of Section \ref{sec:SA_disc} for the fiducial disc model with $\Mp=0.25\Mth$ and surface density slope $p=3/2$. For this particular illustration, we use $\Sigma(R), \Omega(R)$ profiles derived from the corresponding Athena++ simulation at $t = 720 \Pp$ (close in time to Fig. \ref{fig:rec_disc_q15}). For comparison, in Appendix \ref{sec:RWI-another} we provide another illustration of the RWI analysis, this time for a $p=0$ disc.


\subsubsection{Typical mode structure}
\label{sec:RWIlincomp}

\begin{figure}
\centering
\includegraphics[width=0.49\textwidth]{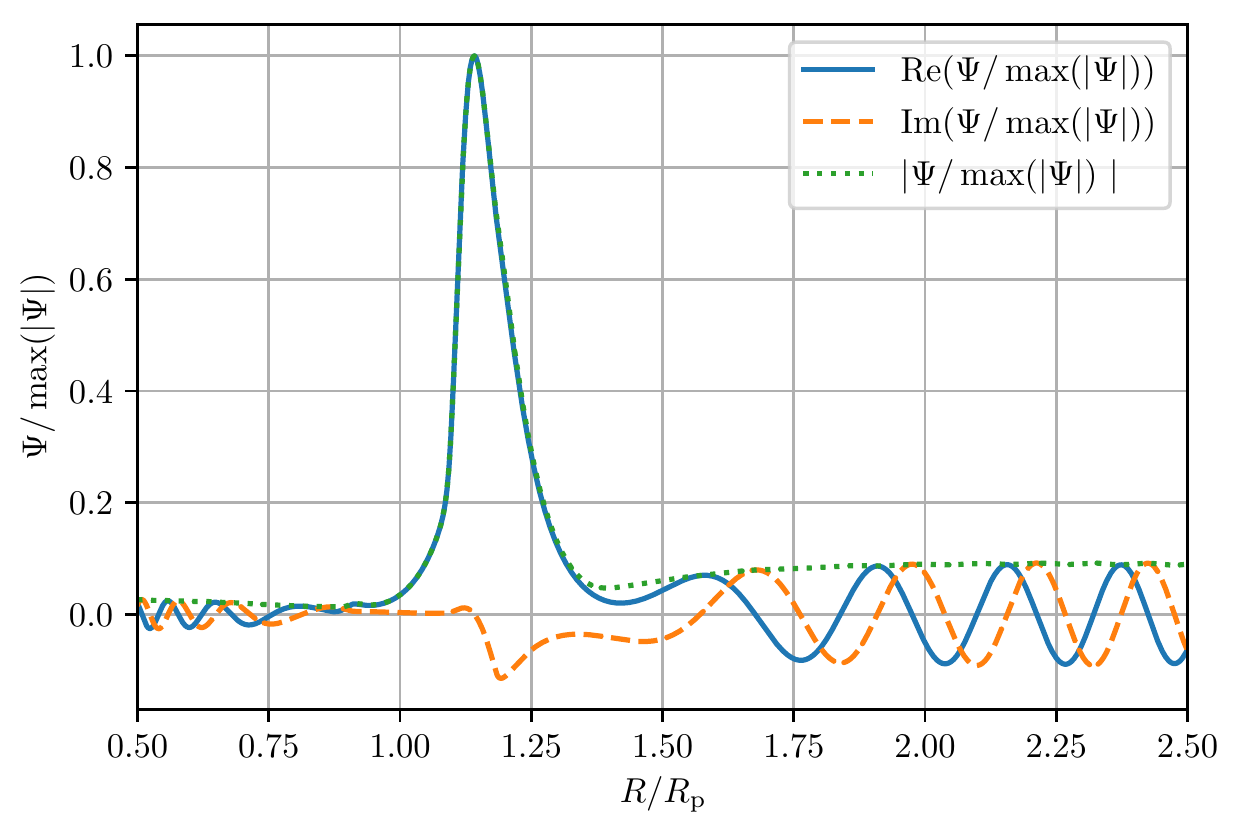}
\includegraphics[width=0.49\textwidth]{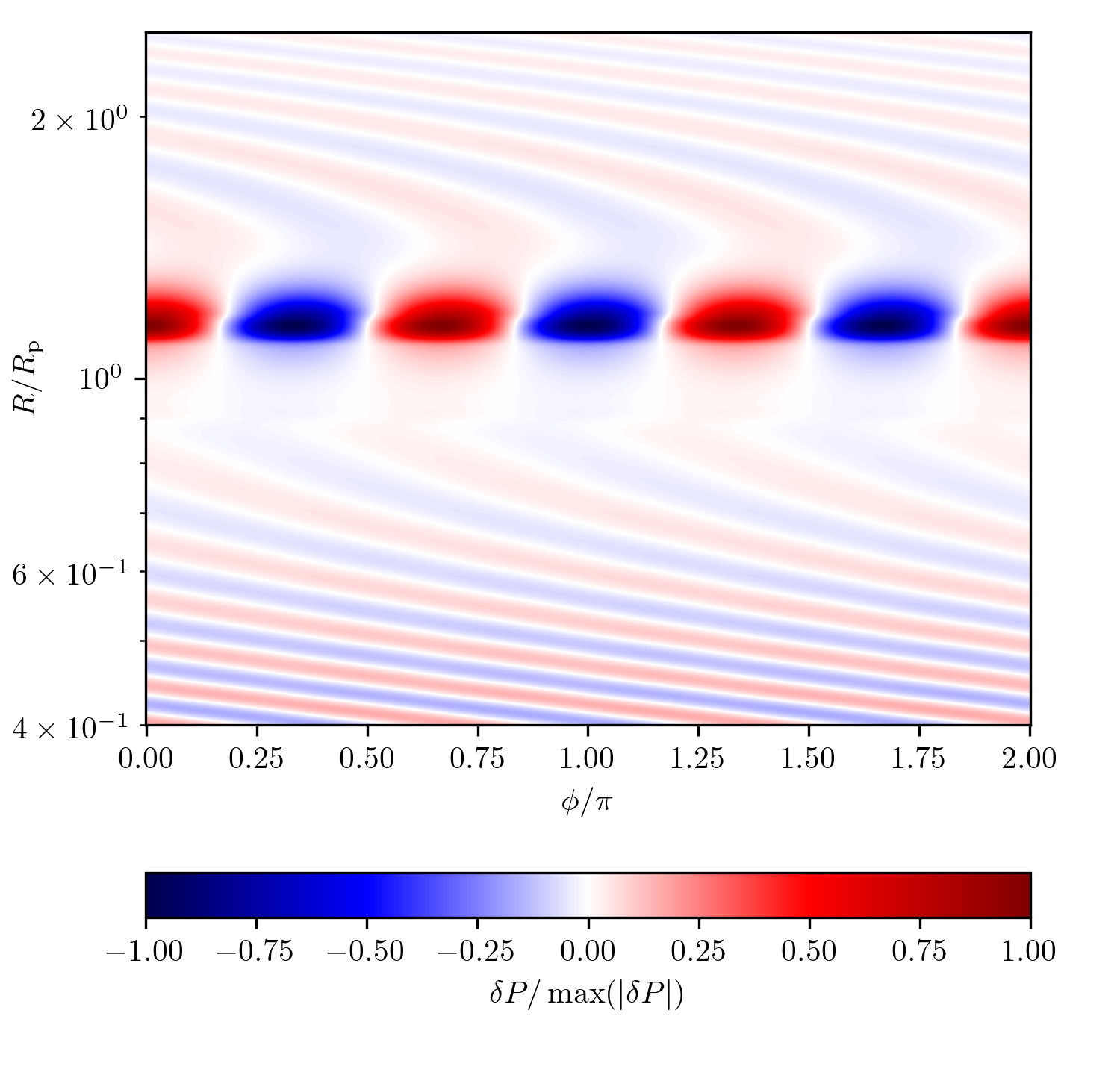}
\vspace*{-.8cm}
\caption{Top: Radial profile of the (normalized) unstable enthalpy eigenmode for $m=3$, associated with the outer gap edge, for the fiducial disc simulation at
$t = 720 \Pp$. We show the real part, imaginary part and absolute magnitude of $\Psi_3$. The frequency of this mode is $\omega = 2.3374+0.01653 \I$. Bottom: Two-dimensional map of the corresponding pressure perturbation.
}
\label{fig:m3_sim_mapdP}
\end{figure}

We start by presenting the typical radial structure of RWI unstable modes,
obtained from linear stability analysis of the fiducial Athena++ model.
In Fig. \ref{fig:m3_sim_mapdP} we show the radial profile of the normalized enthalpy perturbation $\Psi_m/ \max \abs{\Psi_m}$ for $m=3$, since this $m$ dominates the RWI in the fiducial run, see Fig. \ref{fig:dvort_map}. We plot the (normalized) real part, imaginary part and the absolute magnitude of $\Psi_3$ in the top panel. The phase is arbitrarily set such that $\IM (\Psi_m) = 0$ at the location of $\max \vert \Psi_m \vert$\footnote{This corresponds to an azimuthal rotation, under which the axisymmetric problem is invariant.}. One can see that $\Psi_3$ strongly peaks at the vortensity minimum (as expected for trapped modes based on previous studies, e.g. \citealt{Pap1989}, \citealt{Li2000}, \citealt{Lin2010}) at the {\it outer} gap edge, which becomes unstable to RWI earlier than the inner edge. This is expected for $p=3/2$ disc, see \citetalias{Cimerman2021} for details. We also find that spacial trapping of the modes depends on their azimuthal wavenumber $m$, with higher-$m$ modes being less radially confined near the vortensity minimum.

The $m=3$ mode is close to corotating with the background flow ($\omega_R \simeq m \Omega(R_0)$) at the vortensity minimum at $R_0 \simeq 1.17 \Rp$ (see Fig. \ref{fig:rec_disc_q15}a) and has a growth rate corresponding to an e-folding time of about 10 orbits, indicating growth on dynamical timescales. To give an example of the 2D mode structure, we show the $\phi-R$ map of the corresponding normalized pressure perturbation $\delta  P_m/ \max \abs{\delta  P_m}$ in the bottom panel of Fig. \ref{fig:m3_sim_mapdP}. The pressure perturbations ($\delta P_m = \Sigma_0 \Psi_m$) show $m=3$ regions of higher (red) and lower (blue) pressure at the outer gap edge, characteristic of vortices.


\subsubsection{Time-dependence of growth rates and most unstable modes}
\label{subsec:timedepmode}

\begin{figure}
\centering
\includegraphics[width=0.49\textwidth]{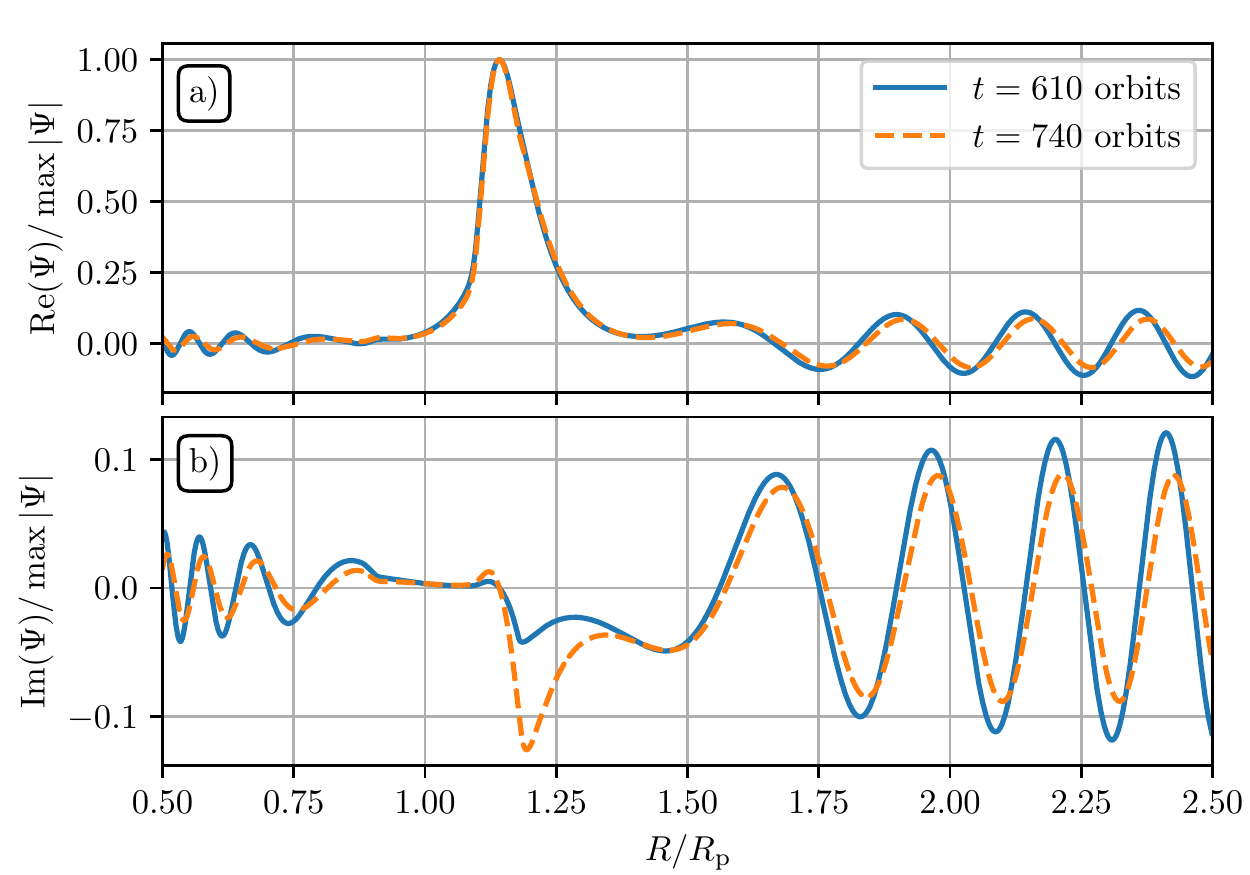}
\vspace*{-0.6cm}
\caption{Radial profile of linearly unstable enthalpy eigenmode $\Psi_3(R)$ for $m=3$, associated with the outer gap edge, for the fiducial disc model at $t = 610 \Pp$ (blue solid) and $740 \Pp$ (orange dashed). One can see that until the instability develops into the non-linear phase, the localized part of the
unstable mode does not change significantly, consistent with the mode growing over long timescales while $\gamma$ changes.
}
\label{fig:m3_time}
\end{figure}

\begin{figure}
\centering
\includegraphics[width=0.49\textwidth]{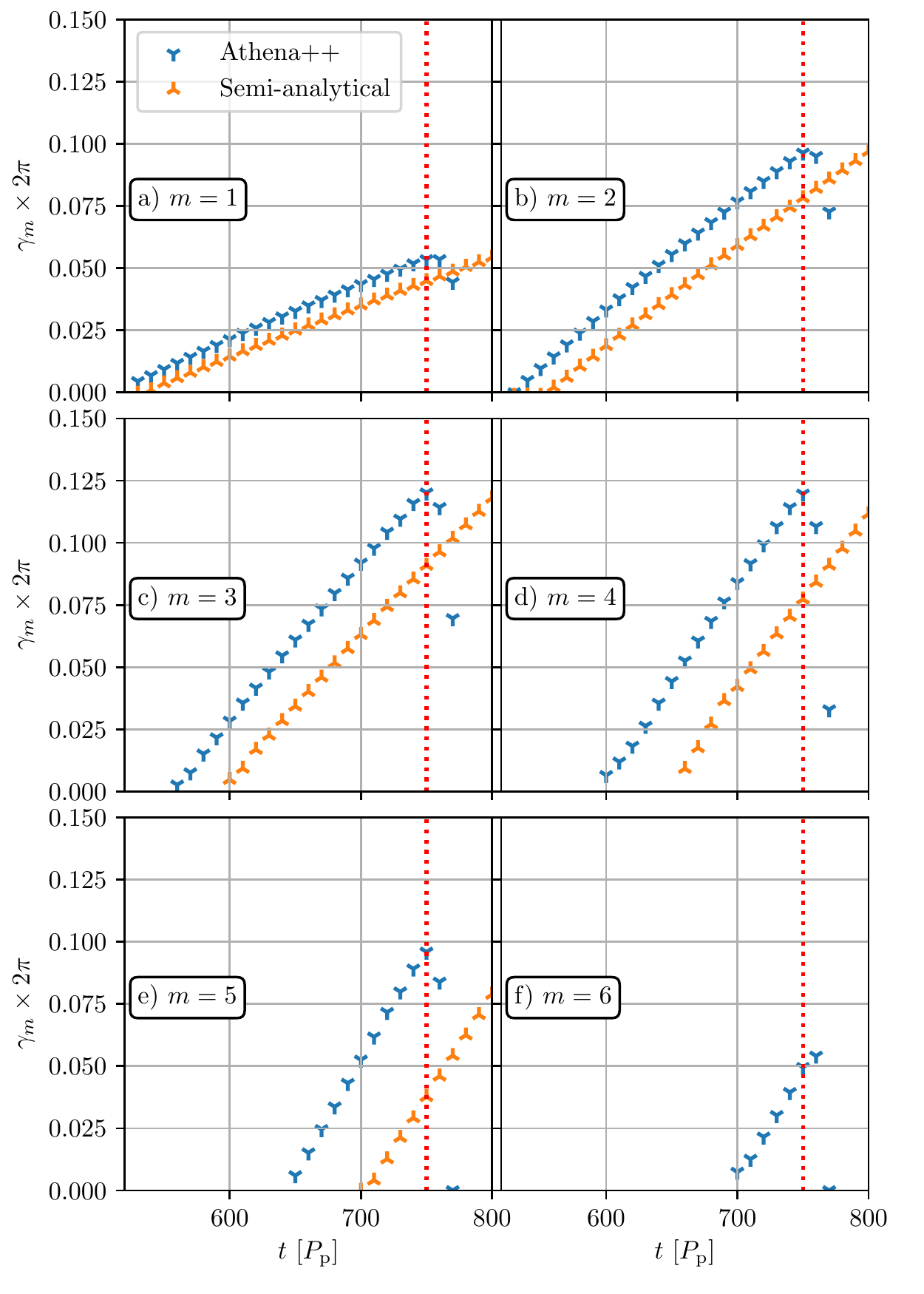}
\vspace*{-0.6cm}
\caption{Instantaneous growth rates $\gamma_m$ of linearly unstable modes, associated with the outer gap edge, found for different $m$ at different times for the fiducial disc and $\Mp/\Mth = 0.25$. We compare results of linear RWI analysis (Section \ref{sec:RWI_eqs}) using disc profiles obtained directly from simulations (blue, downward tripods) and reconstructed using our semi-analytical framework in Section \ref{sec:SA_disc} (orange, upward tripod). The vertical red dotted line marks $t=750$ orbits, when growth rates begin falling off. At this time we observe clearly $m=3$ vortices in the simulation, marking the transition into the non-linear regime of the instability.}
\label{fig:RWI_growthrates_q15}
\end{figure}

As the disc evolves under the action of the planetary perturbation, so do the inputs $\Sigma(R), \Omega(R)$ for the modal equation (\ref{eq:RWI}). As a result, the RWI eigenmodes and, most importantly, eigenfrequencies change in time.

Evolution of the eigenmodes is illustrated in Fig. \ref{fig:m3_time}, where we show the radial structure of $m=3$ mode at $t = 610 \Pp$ (blue, solid) and $t = 740 \Pp$ (orange, dashed) in our fiducial setup with $\Mp= 0.25 \Mth$. One can see that the mode structure at different times is rather similar, with the largest differences in $ \IM(\Psi_3)$ and considerably smaller changes in $ \RE(\Psi_3)$. We can also see that the radial behaviour of the mode phase changes only slightly. These results imply that the {\it radial structure} of the unstable modes change rather slowly during the time in which the instability develops, a property that will later be used in Section \ref{sec:diag}. We find similar results for all other $m$ we consider.

On the other hand, the {\it growth rate} of the mode changes significantly during this interval of time. Indeed, the corresponding eigenfrequencies are $\omega_3(t=610\Pp) = 2.352+0.00566\I$ and $\omega_3(t=740\Pp) = 2.335+0.0184\I$, indicating exponential growth over roughly 28 and 9 planet orbits at these moments of time. As $\vert \RE(\omega) \vert \gg \vert \IM(\omega) \vert = \gamma$, the relative changes of the growth rates are much larger than those of $\vert \RE(\omega) \vert$. This is typical for all linearly unstable RWI modes.

Figure \ref{fig:RWI_growthrates_q15} provides a more detailed illustration of the rapid evolution of the growth rates obtained via our RWI linear stability analysis, and does this for a number of RWI modes ($m = 1$ to $m=6$ in different panels). Orange upward tripods show the growth rates obtained by using $\Sigma(R)$, $\Omega(R)$ given by our semi-analytical reconstruction (Section \ref{sec:SA_disc}) in equation (\ref{eq:RWI}). Blue downward tripods show growth rates obtained when $\Sigma(R)$, $\Omega(R)$ are taken directly from our fiducial simulation. Growth rates are computed every 10 orbits; note that the horizontal axis does not start at $t=0$, but at the moment when unstable modes first appear (for $m=1$).

For both methods of computing the growth rates, unstable modes are found earliest for the lowest azimuthal wave number $m=1$ (ground state). For all $m$, growth rates increase close to linearly with time up until the non-linear phase of instability begins.
This is not surprising since the gap opening process continues and the disc becomes more and more unstable \citep{dvB2007}. The slope of $\gamma_m(t)$ increases with $m$, most notably from $m=1$ to $m=3$. This means that the \textit{instantaneously most unstable} mode will change in time, with higher-$m$ modes with higher $\gamma$ dominating at later time. For all $m$, we find an offset in time between $\gamma_m(t)$ obtained by the two methods $t_\mathrm{off} \simeq 30 \Pp$, with the semi-analytical disc becoming unstable later. This offset does not depend strongly on $m$ and the slope of $\gamma_m(t)$ is comparable between the two methods. These results indicate that the semi-analytical method captures the essential physics and is well suited to determine the timescale for development of the RWI.

Note that the blue tripods, corresponding to $\Sigma$, $\Omega$ derived from a simulation, begin falling off at around $t = 750 \Pp$, which is when strong vortices have developed and the instability transitions into the non-linear regime. This is because perturbations are then strong enough to make the azimuthally averaged disc structure more stable against the RWI\footnote{We note that linear stability analysis is not a suitable method at this stage due to the large amplitude of perturbations.}. On the other hand, the growth rates obtained using the semi-analytical reconstruction do not show such a break and keep increasing monotonically. This is to be expected since there is no instability feedback built into our simple model that would stop $\delta\zeta$ from steadily growing near the planet.


\section{Definition: characteristic timescales of the RWI}
\label{sec:diag}


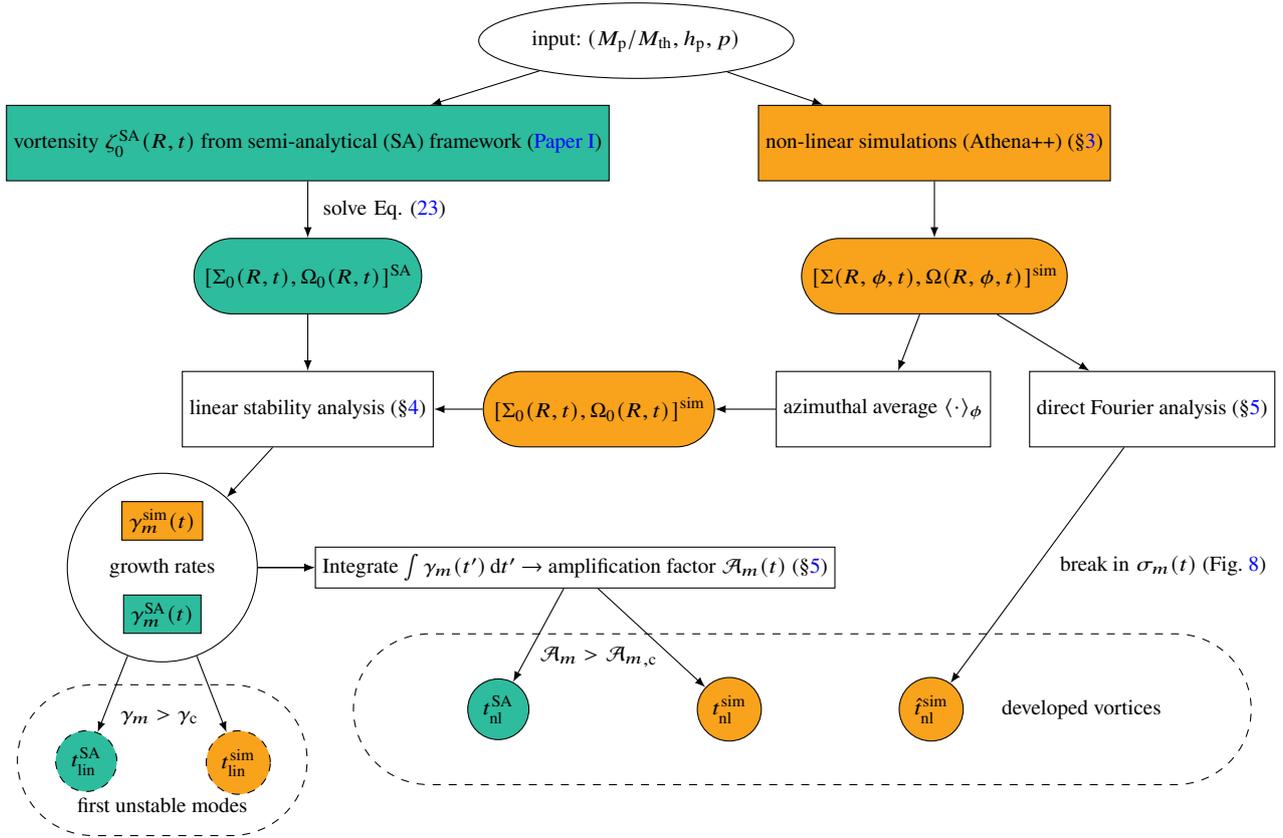
\begin{figure*}
\begin{tikzpicture}

\node[draw,
    ellipse,
    minimum height=1cm,
	 ] (block0) {input: $(\Mp/\Mth, \hp, p)$};

\node[below=0.75 of block0] (bin) {};

\node[draw,
    rectangle,
    minimum width=2.5cm,
    minimum height=1cm,
    left=.25cm of bin,
    fill=SeaGreen] (block1) {vortensity $\zeta_0^\mathrm{SA}(R,t)$ from semi-analytical (SA) framework \citepalias{Cimerman2021}};
    

\node[draw,
    rectangle,
    right=1.5cm of bin,
    minimum width=2.5cm,
    fill=YellowOrange,
    minimum height=1cm,] (block2) {non-linear simulations (Athena++) (\S \ref{sec:hydro-sim})};
    
\node[draw,
    rounded rectangle,
    below=0.75cm of block2,
    minimum width=1.5cm,
    fill=YellowOrange,
    minimum height=1cm,] (block12) {$\left[ \Sigma (R,\phi,t), \Omega (R,\phi,t) \right]^\mathrm{sim}$};

\node[draw,
    rounded rectangle,
    below=.75cm of block1,
    minimum width=1.5cm,
    minimum height=1cm,
    fill=SeaGreen] (block3) {$\left[\Sigma_0(R,t), \Omega_0(R,t) \right]^\mathrm{SA}$};

\node[draw,
    rectangle,
    below=.75cm of block3,
    minimum width=1.5cm,
    fill=white,
    minimum height=1cm,] (block5) {linear stability analysis (\S \ref{sec:RWIlin})};

\node[draw,
    rectangle,
    right=4.5cm of block5,
    minimum width=1.5cm,
    fill=white,
    minimum height=1cm,] (block4) {azimuthal average $\langle \cdot \rangle_\phi$};
        
\node[draw,
    rounded rectangle,
    left=0.8cm of block4,
    minimum width=1.5cm,
	fill=YellowOrange,
    minimum height=1cm,] (block11) {$\left[\Sigma_0(R,t), \Omega_0(R,t) \right]^\mathrm{sim}$};
   
\node[draw,
    rectangle,
    right=.5cm of block4,
    minimum width=1.5cm,
    fill=white,
    minimum height=1cm,] (block6) {direct Fourier analysis (\S \ref{sec:diag})};

\node[rounded rectangle,draw,xshift=4cm,
    minimum width = 2cm, 
    minimum height = 2.5cm,
    left=3.cm of block5,
    yshift=-2.1cm] (blin) {growth rates};

\node[draw,
	  below=-.9cm of blin,
	  fill=SeaGreen
	 ] (block14) {$\gamma_m^\mathrm{SA}(t)$};

\node[draw,
	  above=-.9cm of blin,
	  fill=YellowOrange
	 ] (block15) {$\gamma_m^\mathrm{sim}(t)$};

\node[draw,
	  rectangle,
	  right=.75cm of blin,
	  fill=white
	 ] (block18) {Integrate $\int \gamma_m (t') \,\de t'$ $\rightarrow$ amplification factor $\Am (t)$ (\S \ref{sec:diag})};

\node[below=1.5cm of block18,
	 ] (bint) {};
	 
\node[below=2.cm of bint,
	 ] (bAm) {};

%

\node[draw,
	  circle,
	  left=.5cm of bint,
	  fill=SeaGreen
	 ] (block8) {$\tlsa$};
	 
\node[draw,
	  dashed,
	  circle,
	  below=.9cm of blin,
	  xshift=-1cm,
	  fill=SeaGreen
	 ] (block9) {$\tlrsa$};
	 
\node[draw,
	  dashed,
	  circle,
	  below=.9cm of blin,
	  xshift=1cm,
	  fill=YellowOrange
	 ] (tlrsim) {$\tlrsim$};
	 
\node[below=1.7cm of blin,
	 ] (labelaV) {first unstable modes};
	 
\node[draw,
	  circle,
	  right=1.5cm of bint,
	  fill=YellowOrange
	 ] (block10) {$\tlsim$};
	 
\node[draw,
	  right=1.8cm of block10,
	  fill=YellowOrange,
  	  circle
	 ] (block7) {$\thsim$};

\node[right=.4cm of block7,
	 ] (labelV) {developed vortices};

\node[rounded rectangle,draw,xshift=4cm,
	dashed,
    minimum width = 12cm, 
    minimum height = 2cm] (r) at (block8) {};

\node[rounded rectangle,draw,yshift=.6cm,
	dashed,
    minimum width = 4cm, 
    minimum height = 2cm] (r2) at (labelaV) {};

\draw [arrow] (block0) -- (block1);
\draw [arrow] (block0) -- (block2);

\draw [arrow] (block2) -- (block12);
\draw [arrow] (block12) -- (block4);
\draw [arrow] (block1) -- node [text width=2.5cm,midway,right=.1cm ] {solve Eq. (\ref{eq:ODEsigma})} (block3);
\draw [arrow] (block3) -- (block5);
\draw [arrow] (block4) -- (block11);
\draw [arrow] (block11) -- (block5);
\draw [arrow] (block12) -- (block6);
\draw [arrow] (block6) -- node [midway,right=.2cm ] {break in $\sigma_m (t)$ (Fig. \ref{fig:RWI_growth})} (block7);
\draw [arrow] (blin) -- node [midway,right,yshift=-.3cm] {$\gamma_m > \gamma_\mathrm{c}$} (block9);
\draw [arrow] (blin) -- (tlrsim);
\draw [arrow] (block18) -- (block8);
\draw [arrow] (blin) -- (block18);
\draw [arrow] (blin) -- (block18);
\draw [arrow] (block18) -- node [midway,xshift=-.7cm,yshift=-.25cm] {$\Am > {\Am}_{,\mathrm{c}}$} (block10);
\draw [arrow] (block5) -- (blin);

\end{tikzpicture}
\vspace*{-.5cm}
\caption{Flow chart showing the methods used in our study (rectangles) and their outputs (rounded rectangles and circles). Teal fill marks purely semi-analytical methods, orange fill marks full simulations and no fill marks general methods. The timescales as final outputs are marked by circles. The times when linear stability analysis first predicts unstable modes are $\tlrsa$ and $\tlrsim$, marked by the dashed circles on the left and grouped by the dashed rounded rectangle on the left.
The three timescales characterizing substantial RWI growth --- $t_\mathrm{nl}^\mathrm{SA}$, $\tlsim$, $\thsim$ --- are grouped in the dashed rounded rectangle on the right. See Section \ref{sec:timescales} for details.
}
\label{fig:flowchart}
\end{figure*}

Determination of the timescale for the emergence of vortices at the edges of planet-induced gaps is the main goal of this paper. To that end, we now define a number of important timescales that characterize different stages of the vortex development via the RWI. Given the variety of methods that we employ to study this problem, some of the timescales also depend on the particular method used in defining them, see Fig. \ref{fig:flowchart} for illustration. In particular, we use the superscript 'SA' for the timescales based on the semi-analytical disc reconstruction technique of Section \ref{sec:SA_disc}, while the ones utilizing the simulation data in any way are denoted 'sim'.  We also use different timescales to characterize the onset and non-linear development of the instability, as described below.


\subsection{Timescales based on linear RWI stability analysis}
\label{sec:RWI-time}


As the planet opens a deeper and deeper gap, the gap edge will at some point become unstable, which results in RWI modes computed by our linear analysis attaining $\gamma_m > 0$. This allows us to define several characteristic timescales, as follows.\\


\noindent{\bf $\tlrsa$: onset of instability with semi-analytic reconstruction}\\
We define $\tlrsa$ as the time after which linear stability analysis first predicts unstable RWI modes with a growth rate greater than a critical value $\gamma_m > \gamma_\mathrm{crit} = 10^{-2}/2 \pi$ (corresponding to exponential growth with an e-folding time of 100 planet orbits). This particular definition assumes that the radial structure of the disc used in our linear stability analysis has been obtained through the semi-analytical reconstruction described in Section \ref{sec:SA_disc}.\\

\noindent {\bf $\tlrsim$: onset of instability with simulation input}\\
The time $\tlrsim$ is defined analogous to $\tlrsa$ but using azimuthal averages of 2D simulation results as $\Sigma(R), \Omega(R)$ inputs for the RWI linear stability analysis.\\

\begin{figure}
\centering
\includegraphics[width=0.49\textwidth]{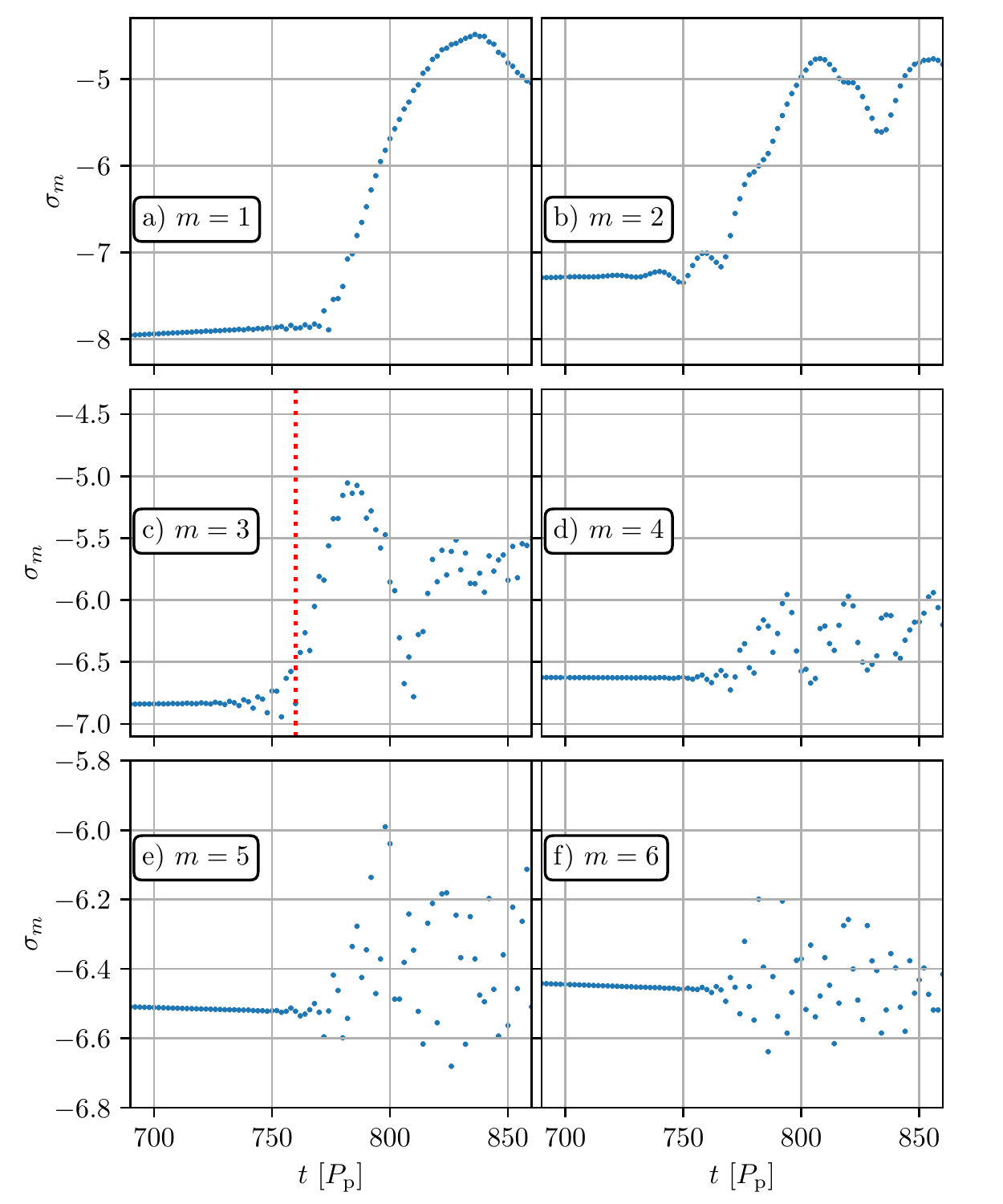}
\vspace*{-.4cm}
\caption{Logarithmic measure of the relative amplitudes of the Fourier components of the surface density perturbation $\vert \Sigma_m \vert / \Sigma$ measured in our runs, integrated over a radial range around the outer gap edge ($\sigma_m$, see equation \ref{eq:sigmam}) and shown as a function of time for different $m$. The break in the behaviour of $\sigma_m$ indicates the transition to the non-linear phase of RWI in our simulations.}
\label{fig:RWI_growth}
\end{figure}

\noindent {\bf $t_\mathrm{nl}^\mathrm{SA}$: developed RWI with semi-analytic reconstruction}\\
While the above procedure predicts when linear instability \textit{first} sets in, it does not tell us when the instability has amplified to a point where substantial vortices appear.

Results of Section \ref{subsec:timedepmode} indicate that the radial structure of an unstable mode does not vary much as the instability develops, while the growth rate changes substantially. If we take two moments of time during this phase, $t_\mathrm{i}$ and $t_\mathrm{f}$, then the mode amplitudes at these times are related by the integral of instantaneous growth-rates over time:
\begin{align}
	\ln \frac{\abs{\delta  \Sigma_m(R,t_\mathrm{f})}}{\abs{\delta  \Sigma_m(R,t_\mathrm{i})}}
	= \int_{t_\mathrm{i}}^{t_\mathrm{f}} \gamma_m \,\de t \equiv \Gamma_m.
\label{eq:amp_rel}
\end{align}
In other words,
\begin{align}
	\abs{\delta  \Sigma_m(R,t_\mathrm{f})} = \exp \left( \Gamma_m \right) \abs{\delta  \Sigma_m(R,t_\mathrm{i})} = \Am \abs{\delta  \Sigma_m(R,t_\mathrm{i})},
\end{align}
where we have defined the amplification factor $\Am \equiv \exp \left( \Gamma_m \right)$.

Using this logic, we can define a timescale $t_\mathrm{nl}^\mathrm{SA}$ as the time over which an initial perturbation of order $\vert \delta  \Sigma \vert /\Sigma \simeq \mathcal{A}_{m,\mathrm{crit}}^{-1}$ grows to order unity:
\begin{align}
	\Am(\tn) \geq \mathcal{A}_{m,\mathrm{crit}},
\end{align}
where $\mathcal{A}_{m,\mathrm{crit}}$ is some critical value (see next) and the growth rates in (\ref{eq:amp_rel}) are computed using the inputs based on semi-analytical reconstruction. By definition, we always have $\tlrsa < \tlsa$. 

By comparing $\tlsa$ with the time when non-linear simulations show substantial RWI vortices (see Section \ref{sec:thsim}), we find good agreement when setting $\mathcal{A}_{m,\mathrm{crit}}=10^4$, which gives an estimate of initial deviations from axisymmetry ($\delta \Sigma/\Sigma \simeq 10^{-4}$). Due to the exponential growth of the instability, the timescale is not very sensitive to this choice. Here we assume that the initial noise is of similar amplitude for all mode numbers. In real systems, there might be preferential forcing of specific modes. This might lead to a different offset between $\tlrsa$ and $\tlsa$ as $m$ varies.\\

\noindent {\bf $\tlsim$: developed RWI with simulation inputs}\\
We also define another time $\tlsim$ for developed RWI, analogous to $\tlsa$ but now using the azimuthally-averaged simulation data when computing the growth rates $\gamma_m$, which enter the equation (\ref{eq:amp_rel}).


\subsection{$\thsim$: simulation-only timescale}
\label{sec:thsim}


Finally, we also diagnose the non-linear development of the instability using a time $\thsim$ defined using {\it only} the simulation results, i.e. not resorting to the RWI linear stability analysis as we do for all other characteristic times defined in Section \ref{sec:RWI-time}. This is done by tracking the non-axisymmetric density structure in the simulations, the amplitude of which is interpreted as indicating the RWI development.

In practice, we first perform Fourier decomposition of the surface density perturbation of our 2D simulation data:
\begin{align}
	\Sigma_m (R,t) \equiv  \frac{1}{2 \pi} \int_0^{2 \pi} \Sigma(R,\phi,t) \exp (-\I m \phi) \,\de \phi,
\end{align}
with the prefactor chosen such that $\azav{\Sigma} = \Sigma_0$. We then define the quantity
\begin{align}
	\sigma_m \equiv \ln \left( \int_{R_\mathrm{l}}^{R_\mathrm{u}} \frac{\abs{ \Sigma_m}}{\Sigma _0} \,R \de R \right),
	\label{eq:sigmam}
\end{align}
where we perform the integral over a radial region $R_\mathrm{l} < R < R_\mathrm{u}$ around the gap edge, where RWI unstable modes peak in amplitude; at the outer gap edge we typically take $R_\mathrm{l} = 1.05 \Rp$ and $R_\mathrm{u} = 1.5 \Rp$. A similar metric has been used for example by \citet{Les2015}.

Once the instability develops, we expect $\sigma_m$ to become non-zero and to grow linearly in time allowing us to diagnose the onset of RWI. However, there is a complication: due to the non-axisymmetric perturbations induced by the planet, $\abs{\Sigma_m} \neq 0$ at all times, even before the gap edge becomes unstable. This is illustrated in Fig.  \ref{fig:RWI_growth}, where we plot $\sigma_m(t)$ for the fiducial simulation for $ 1 \leq m \leq 8$. One can see that, indeed, $\sigma_m$ is always non-zero and shows slow variation early on. The weak time dependence is caused by the effect of gap opening on the planet-induced density waves\footnote{The radial locations of Lindblad resonances depend on their order, such that local changes to the relative surface density will affect the excitation of modes with different $m$ differently \citep[e.g.][]{Petrovich2012}.}.

Nevertheless, we can still associate the onset of instability with the {\it break} in the behaviour of $\sigma_m(t)$, which is clearly present in all panels. Indeed, beginning at around $t = 720 \Pp$, small oscillations occur for the three ($1 \leq m \leq 3$) lowest mode numbers, which are most pronounced for $m=2$ and $m=3$. Around $t= 760 \Pp$, there is a clear break in $\sigma_m(t)$ behaviour of the $m=3$ mode (red dotted line vertical in panel c), shortly after which all $1 \leq m \leq 3$ show strong growth\footnote{Modes with $m \geq 4$ show more noisy behaviour of $\sigma_m(t)$ instead of coherent growth.} of $\sigma_m(t)$, indicating (close to exponential) growth of $\Sigma_m$ for around 50 orbits, after which $\sigma_m$ saturates.
The break thus corresponds to a time, which we call $\thsim$, when RWI has grown substantially and started dominating\footnote{Note that $\thsim$ occurs slightly earlier than RWI saturation in simulations (which is not very well defined). However the relative difference between the two moments of time is typically $\lesssim 5\%$, so we define $\thsim$ based on the initial break of $\sigma_m(t)$.} over the non-axisymmetric perturbation due to the planet. This interpretation is confirmed by examining Figs. \ref{fig:dvort_map} \& \ref{fig:dvort_azav_map}, which show that  at $t \approx \thsim=760 \Pp$ vortices become visible and comparable in amplitude to planetary perturbations in 2D maps of $\Sigma$ and vortensity perturbations. As the determination of $\thsim$ uses only the simulation data, it allows for an independent verification of our semi-analytical calculations.
\\

To summarize, we can define five different characteristic times, two of which --- $\tlrsa$ and $\tlrsim$ --- describe first appearance of unstable modes, while three others ---  $\tlsa$, $\tlsim$, and $\thsim$ --- mark the development of the non-linear phase of the RWI. These different metrics are compared in the following section.


\section{Results: timescales for instability}
\label{sec:timescales}


\begin{table}
\centering
\caption{Complete list of hydro simulations that were performed. We report planet mass $\Mp$ (in units of $\Mth$), disc parameters ($p$, $\hp$), simulation time (in orbits) at which we observe the RWI at inner and outer gap edges, respectively. Last column indicates which gap edge shows vortices first. Our fiducial simulation is shown in boldface.
}
\label{tab:params}
  \begin{tabular}{lccccc}
    \hline
    $\Mp/\Mth$ & $p$ & $\hp$  & $\hat{t}^\mathrm{sim}_\mathrm{inner} /\Pp$ & $\hat{t}^\mathrm{sim}_\mathrm{outer} / \Pp$ & most unstable gap edge\\
    \hline
	0.1 &  1.5 & 0.05  & $> 6800$ & $> 6800$ & -\\
	0.1 &  1.5 & 0.07  & $> 4200$ & $> 4200$ & -\\
	0.1 &  1.5 & 0.10  & $> 2970$ & $> 2970$ & -\\
	0.1 &  0 & 0.05  &  $> 5850$ & $> 5850$ & -\\
	0.1 &  0 & 0.07  &  $> 4750$ & $> 4750$ & -\\
	0.1 &  0 & 0.10  &  $> 2800$ & $> 2800$ & -\\
    \hline
   	{\bf 0.25} & {\bf 1.5} & {\bf 0.05}  & {\bf 880} & {\bf 760} & {\bf outer} \\
	0.25 &  1.5 & 0.07  & 690& 540 & outer \\
	0.25 &  1.5 & 0.10  & 520& 380 & outer \\
	0.25 &  0 & 0.05  & 710& 1000 & inner \\
	0.25 &  0 & 0.07  & 510& 810 & inner \\
	0.25 &  0 & 0.10  & 330& 450 & inner \\
	\hline
   	0.5 &  1.5 & 0.05  & 164 & 146 & outer \\
	0.5 &  1.5 & 0.07  & 130 & 112 & outer \\
	0.5 &  1.5 & 0.10  & 94 & 82 & outer \\
	0.5 &  0 & 0.05  & 144 & 168 & inner \\
	0.5 &  0 & 0.07  & 98 & 128 & inner \\
	0.5 &  0 & 0.10  & 72 & 112 & inner \\
	\hline
  \end{tabular}
\end{table}

We now describe the results on the various instability timescales (defined in previous section) obtained by applying the methods described in Sections \ref{sec:hydro-sim} \& \ref{sec:RWIlin} to a variety of disc-planet setups. We vary a number of parameters of the problem, choosing three values each for the normalized planet mass\footnote{The lower limit on $(\Mp/\Mth)$ is due to the requirement of increasingly high resolutions to correctly capture vortensity generation by low mass planets (see \citetalias{Cimerman2021}), coupled with increasing timescale for the onset of RWI.} $\Mp/\Mth \in \lbrace 0.1, 0.25, 0.5 \rbrace$, and the disc scale-height $\hp \in \lbrace 0.05, 0.07, 0.1 \rbrace$ and two values for the surface density slope of the background disc $p \in \lbrace 0, 1.5 \rbrace$. Covering all possible combinations, this gives a total of 18 disc models listed in Table \ref{tab:params}. 

For every disc model we run a direct simulation to get the simulation-based timescales $\tlrsim$, $\tlsim$, and $\thsim$, and use the semi-analytical reconstruction of Section \ref{sec:SA_disc} to obtain $\tlrsa$, $\tlsa$. For $\Mp = 0.1\Mth$, simulations could not be run long enough to develop RWI, providing only a lower limit on the instability timescale $\thsim$. To carry out our linear analysis (Section \ref{sec:RWI-time}) and to observe the development of RWI directly (Section \ref{sec:thsim}), we sample the disc state every 10 orbits in runs with $\Mp = 0.25\Mth$, and every 2 orbits in the highest $\Mp = 0.5\Mth$ runs.


\subsection{First appearance of unstable modes}
\label{sec:time-lin}


\begin{figure}
\centering
\includegraphics[width=0.49\textwidth]{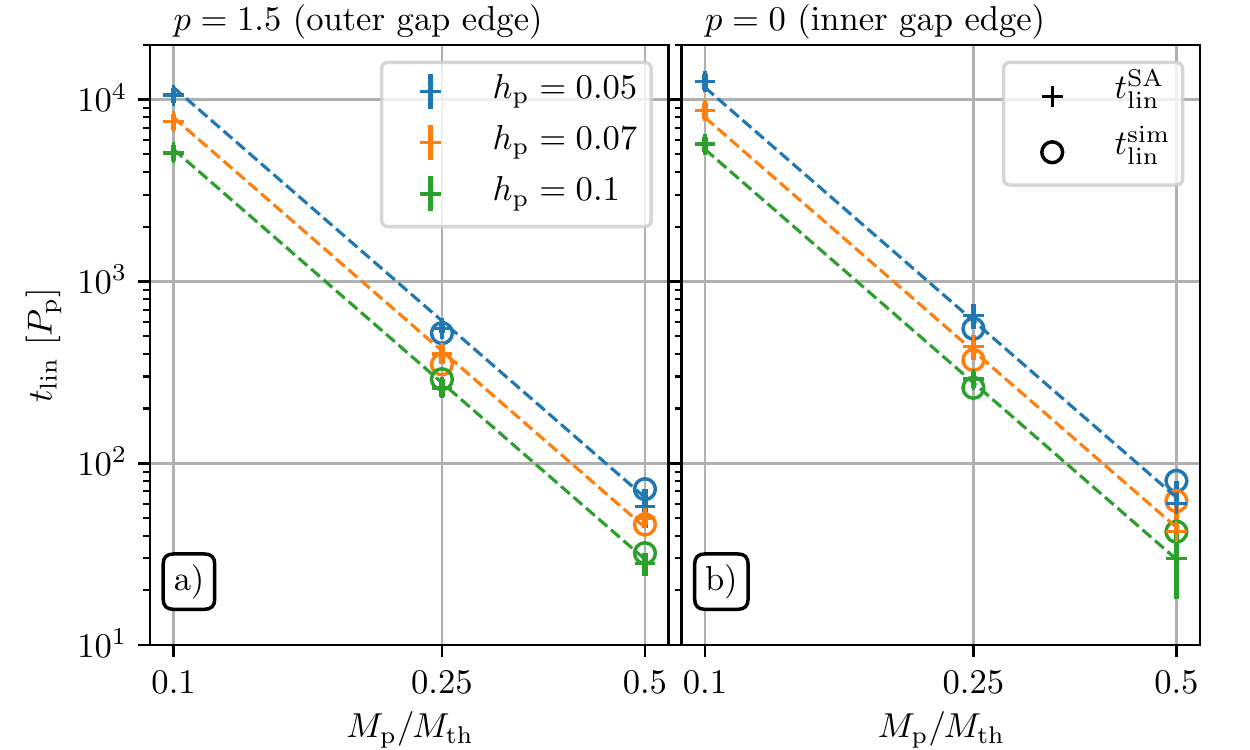}
\vspace*{-.3cm}
\caption{
Characteristic timescales for the appearance of first unstable RWI modes. We show $\tlrsa$ (crosses) for all parameter sets and $\tlrsim$ (circles) where available, see Section \ref{sec:RWI-time} for definitions. We show results for 
(a) the fiducial disc ($p=1.5$) and (b) the constant surface density disc $(p=0)$. Colours correspond to different values of $\hp$. The dashed lines show a power law fit for $\tlrsa$ given by equation (\ref{eq:taufit}) with parameters (\ref{eq:fit-l}).
}
\label{fig:taulin}
\end{figure}

We start by presenting results for $\tlrsa$ and  $\tlrsim$ --- the times when linearly unstable modes with $\gamma_m > \gamma_\mathrm{crit}$ first appear (see Section \ref{sec:RWI-time}) --- allowing us to compare RWI linear analysis with inputs based on simulations and semi-analytical method. In Fig. \ref{fig:taulin}, we plot $\tlrsa$ as crosses for different $\hp$ (colours) as a function of normalized planet mass on a log-log scale. In panels (a) and (b) we consider discs with $p=1.5$ and $p=0$, examining onset of RWI at the outer and inner gap edge, respectively.

For all parameter sets except those including $\Mp = 0.1 \Mth$, we also show $\tlrsim$ (circles). For the lowest planet mass explored, $\Mp = 0.1 \Mth$, simulations do not show RWI vortices at the end of run time indicated in Table \ref{tab:params}, e.g. after $t_\mathrm{end} = 6800 \Pp$ for the disc with fiducial parameters\footnote{This run required more than half a year of continuous computation.}. This is in full agreement with the estimates of $\tlrsa$ for these runs, which all exceed the run times, see Fig. \ref{fig:taulin}. 

We can make several observations based on this figure. First, there is generally good agreement between $\tlrsa$ and  $\tlrsim$ for any given disc-planet model, confirming the robustness of our semi-analytic reconstruction technique. The agreement is slightly worse for models with $p=0$, but this is to be expected because of the secondary arm formation in the inner disc, see Appendix \ref{sec:RWI-another}.

Second, for both values of $p$, the behaviour of $\tlrsa$ and  $\tlrsim$ in Fig. \ref{fig:taulin} suggests a power law fit for the time $\tlr$ when the first unstable modes appear of the form (we will use this fit also in the next section for $\tn$)
\begin{align}
\{\tlr,\tn\} = A \Pp \left( \frac{\Mp}{\Mth} \right)^\alpha \hp^\beta
\label{eq:taufit}
\end{align}
or, equivalently,
\begin{align}
\log \frac{\{\tlr,\tn\}}{\Pp} = \log A + \alpha \log \frac{\Mp}{\Mth} + \beta \log \hp.
\label{eq:pl_t}
\end{align}
To obtain the parameters of this fit, we perform a least-squares regression of $\tlrsa$ data (as we have them available for all $\Mp$, even the lowest-mass case $\Mp=0.1\Mth$) in log-log space over all 18 data points, including $p=1.5$ and $p=0$. Our resultant fit of the form (\ref{eq:taufit}) is shown via dashed lines in Fig. \ref{fig:taulin} and has parameters
\begin{align}
(\log A,\alpha,\beta)_{\tlr} = (-0.63 \pm 0.09, -3.23 \pm 0.05, -1.13 \pm 0.08).
  \label{eq:fit-l}
\end{align}
so that $A\approx 0.23$. The maximum relative deviation between this fit and the $\tlrsa$ data is 11\% over all points, and we see a good match for both surface density slopes. 

The power law indices indicate that the planet mass and disc scale-height have the strongest influence on $\tlr$, with the former dominating, and the surface density slope having insignificant effect. 


\subsection{Timescale for developed RWI (vortex formation)}
\label{subsec:tauhat}


\begin{figure}
\centering
\includegraphics[width=0.49\textwidth]{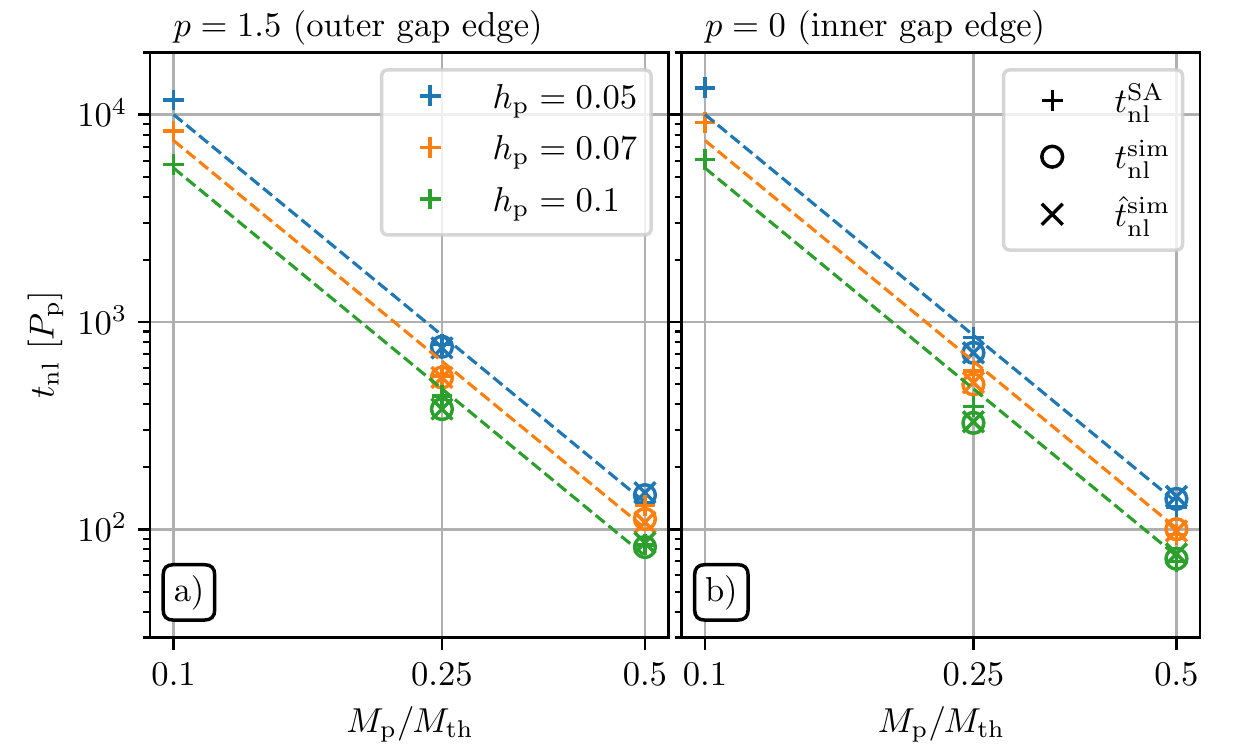}
\vspace*{-.3cm}
\caption{
Timescales characterizing the non-linear development of RWI: $\tlsa$ (pluses), $\tlsim$ (circles), and $\thsim$ (crosses), see Sections \ref{sec:RWI-time},\ref{sec:thsim} for definitions. Panels again show (a) fiducial disc ($p=1.5$) and (b) constant surface density disc $(p=0)$, colours indicate different $\hp$. The dashed lines show a power law fit (\ref{eq:taufit}) with parameters (\ref{eq:fit-nl}) for all 18 data points for $\tlsa$.}
\label{fig:tauhatlin}
\end{figure}

Perhaps more important than $\tlr$ from a practical standpoint is the time-scale $\tn$ for well-developed RWI, indicating when large-amplitude vortices appear and can be observed directly. We use the three previously defined (Section \ref{sec:diag}) time scales --- $\tlsa$, $\tlsim$, and $\thsim$ --- as proxies for $\tn$ for all parameter sets, except those with $\Mp/\Mth = 0.1$ for which only $\tlsa$ is available, since simulations have not become unstable by the end of the run time.

In Fig. \ref{fig:tauhatlin}, we show these timescales in a way similar to Fig. \ref{fig:taulin}, see the legend. The agreement we find between $\tlsim$ (RWI analysis using simulation data as input) and $\thsim$ (timescale based directly on simulations) is excellent for all simulations, with maximum relative deviations between the timescales of 5\%. This justifies our method of integrating the linear growth rates (see  Section \ref{sec:RWI-time}) as a good way of predicting the emergence of fully-developed vortices. We also find that the semi-analytical estimate $\tlsa$ gives a maximum relative deviation of 15\% compared to $\thsim$, once again, supporting our gap reconstruction technique of Section \ref{sec:SA_disc}. We use this relative deviation as a simple estimate for the uncertainty of $\tlsa$ below.

Again, a least square fit for $\tn$ in the form (\ref{eq:pl_t}) is appropriate, and we obtain its parameters by linear regression of $\tlsa$ data as
\begin{align}
  (\log A,\alpha,\beta)_{\tn} = (0.21 \pm 0.13, -2.68 \pm 0.06, -0.86 \pm 0.11).
  \label{eq:fit-nl}
\end{align}
so that $A\approx 1.6$. With these parameters, equation (\ref{eq:taufit}) yields the vortex emergence timescale in physical units as 
\begin{align}
\tau & \approx 4\times 10^3\mbox{yr}\left(\frac{\Mp}{\Mth}\right)^{-2.7}
    \left(\frac{\Rp}{50\mbox{AU}}\right)^{1.5}
    \left(\frac{\hp}{0.1}\right)^{-0.86}
    \left(\frac{M_\star}{M_\odot}\right)^{-0.5}
\label{eq:T_vrt0}\\
& \approx 3\times 10^4\mbox{yr}\left(\frac{\Mp}{0.5M_\mathrm{J}}\right)^{-2.7}
    \left(\frac{\Rp}{50\mbox{AU}}\right)^{1.5}
    \left(\frac{\hp}{0.1}\right)^{7.2}
    \left(\frac{M_\star}{M_\odot}\right)^{2.2}.
	\label{eq:T_vrt}
\end{align}

In agreement with the results of \citetalias{Cimerman2021}, we find that in a disc with constant surface density ($p=0$), the inner gap edge is more unstable than the outer one, opposite to the findings for the fiducial $p=3/2$ disc. We also see that the RWI stability of the inner gap edge is affected to a certain degree by the formation and shocking of a secondary (and higher order) spiral arms in the inner disc \citep{Bae2018,Miranda2019I}. This effect is not captured by our gap reconstruction technique (Section \ref{sec:SA_disc}), while being fully accounted for in simulations. Nevertheless, our semi-analytical predictions for the RWI development at the inner gap edge (e.g. in $p=0$ discs) match $\thsim$ quite well, see Fig. \ref{fig:tauhatlin}.


\subsection{Theoretical estimate}
\label{sec:theorytau}

\begin{figure}
\centering
\includegraphics[width=0.49\textwidth]{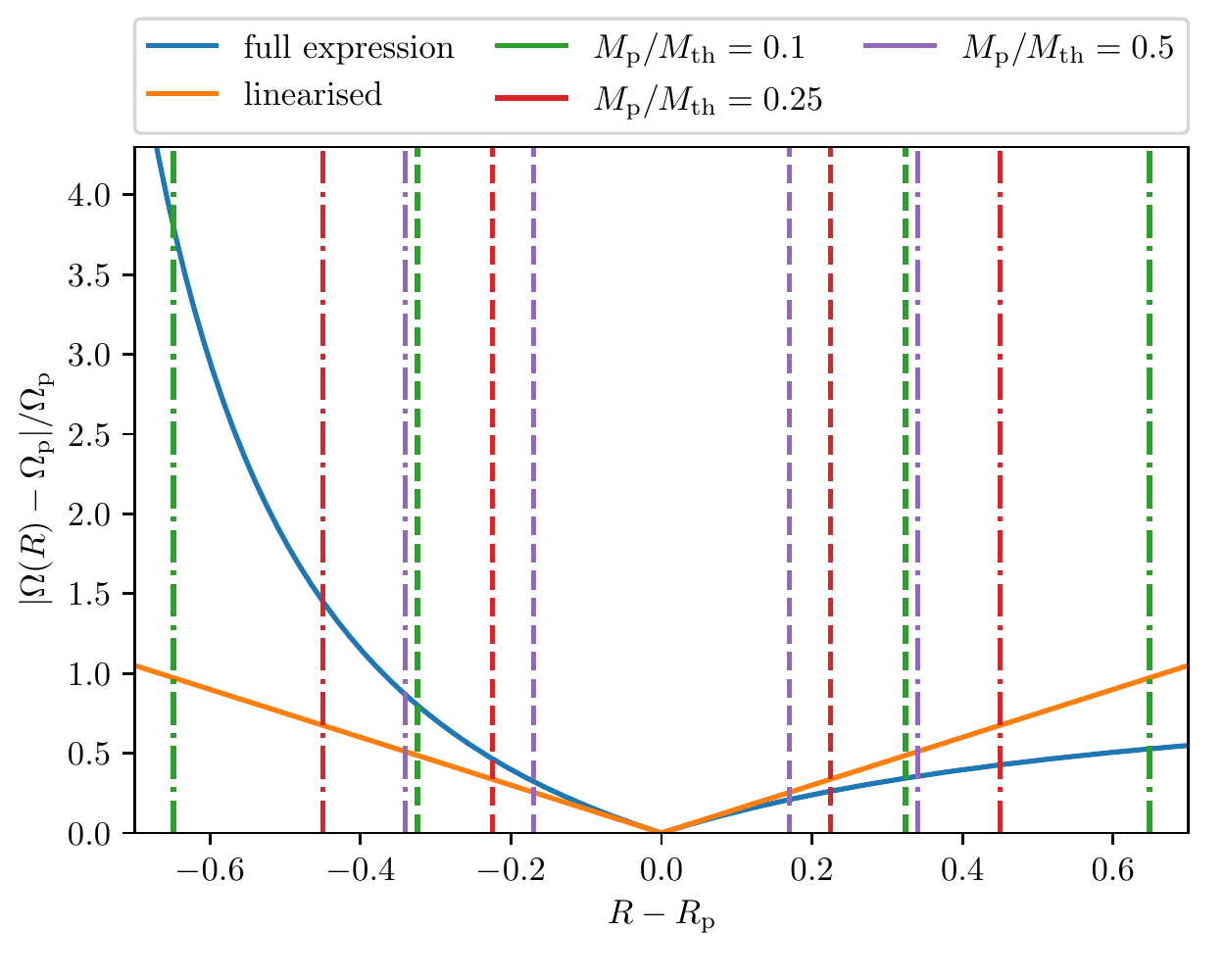}
\vspace*{-.6cm}
\caption{Relative orbital frequency of gas parcels w.r.t. the planet-driven shock. We show the full expression (blue) and its linearised form (equation (\ref{eq:linsyn}), orange). Vertical lines indicate a radial separation $\vert R-\Rp \vert = 3 \lsh$ from the planet for different $\Mp/\Mth$ (colours) and $\hp = 0.05$ and $\hp = 0.1$ (dashed and dash-dotted lines, respectively).
}
\label{fig:lin_syn}
\end{figure}

From our previous discussion (Section \ref{sec:RWI_eqs}) and the results of \citetalias{Cimerman2021} it is natural to expect that the timescale $\tau$ on which the disc structure changes and the RWI sets in  should depend on the vortensity production rate by the planetary shocks $\Ssh$. According to equation (\ref{eq:syn}), $\Ssh$ is given by the product of the vortensity jump at the shock $\Delta\zeta$ and the synodic frequency of a shocked gas parcel w.r.t. the planet. To lowest order, we can estimate the synodic frequency using the local approximation, i.e. expanding it to linear order in $R-\Rp$:
\begin{align}
	\vert \Omega(R) - \Omega_\mathrm{p} \vert \simeq \frac{3 \Omega_\mathrm{p}}{2} \left\vert \frac{R - \Rp}{\Rp} \right\vert.
	\label{eq:linsyn}
\end{align}
Figure \ref{fig:lin_syn} illustrates the performance of this approximation, with the local expansion (\ref{eq:linsyn}) shown in orange being compared to $\vert\Omega-\Omega_\mathrm{p}\vert$ (solid blue). Vertical coloured lines correspond to $R=\Rp\pm 3\lsh$, computed for different values of $\hp$ and $\Mp/\Mth$: we know from \citetalias{Cimerman2021} that $\Delta \zeta$ is significant only at radii satisfying $\lsh \lesssim \vert
R - \Rp \vert \lesssim 3 \lsh$ (for $\Mp\lesssim \Mth$). One can see that within this range the local approximation (\ref{eq:linsyn}) provides a decent fit for small $\lsh$, with order unity deviations becoming particularly noticeable for highest $\hp$ and lowest $\Mp/\Mth$, mainly in the inner disc. This has important implications (see Section \ref{sec:zeta_th}), but for the sake of our simple argument we will adopt the approximation (\ref{eq:linsyn}) in what follows.

Moreover, we have shown in \citetalias{Cimerman2021} that the vortensity jump at the shock $\Delta \zeta\propto (\Mp/\Mth)^3$, but is a weak function of the disc aspect ratio $\hp$. With this in mind, using $\vert R-\Rp\vert\sim\lsh$ and recalling that $\lsh\approx \Rp\hp(\Mp/\Mth)^{-2/5}$ \citep{Goodman2001}, equations (\ref{eq:syn}) and (\ref{eq:linsyn}) yield
\begin{align}
	 \pd{\zeta}{t} =\Ssh\propto \left( \frac{\Mp}{\Mth} \right)^{3} \times \Omega_\mathrm{p}\frac{\lsh}{\Rp} \propto \Omega_\mathrm{p} \hp \left( \frac{\Mp}{\Mth} \right)^{2.6}.
	 \label{eq:dzdt}
\end{align}
Note that $\Ssh$ depends on $\hp$ both explicitly and through $\Mth$, see equation (\ref{eq:Mth}). 

Let us now make a simple assumption that RWI sets in and vortices appear at the time $\tau$ corresponding to the moment when $\delta\zeta$ (e.g. its peak or trough value, ignoring the radial structure for simplicity) reaches some threshold value $\zeta_\mathrm{th}$; we discuss the validity of this assumption next in Section \ref{sec:zeta_th}. In other words, in this approach $\tau$ should be determined from the condition
\begin{align}
\int^\tau_0\Ssh(t)\de t\approx  \zeta_\mathrm{th}.
\label{eq:inst_cond}
\end{align}
With $\Mp$, $\hp$ and other parameters not changing in time, as we assume in this work (but see Section \ref{subsec:appl}), $\Ssh$ remains constant. Then the integral in (\ref{eq:inst_cond}) is trivial, and the time for vortices to appear is simply $\tau\approx \Ssh^{-1}\zeta_\mathrm{th}$. Given the expression (\ref{eq:dzdt}), it then follows that the characteristic timescale $\tau$ for the emergence of vortices at the edges of a planetary gap should scale as
\begin{align}
\tau \propto \Pp \hp^{-1} \left( \frac{\Mp}{\Mth} \right)^{-2.6}.
\label{eq:estdzetadt}
\end{align}

Given the simplicity of our argument, this estimate may apply to both $\tlr$ and $\tn$. Regarding the former, equation (\ref{eq:fit-l}) implies stronger $\tlr$ dependence on planet mass, $\tlr \appropto \left( \Mp / \Mth \right)^{-3.2}$, than in (\ref{eq:estdzetadt}). On the other hand, the value of $\alpha$ in the $\tn$ fit (\ref{eq:fit-nl}) is close to $-2.6$ in equation (\ref{eq:estdzetadt}). This improved agreement for $\tn$ is likely a coincidence, since calculation of $\tlsa$ (leading to $\tn$) takes into account the evolution of $\gamma_m(t)$ during the linear stage of instability, while the estimate (\ref{eq:estdzetadt}) ignores such details.

Note also that our estimate (\ref{eq:estdzetadt}) reproduces reasonably well the explicit dependence of the instability timescale on $\hp$. Indeed, $\tlr \appropto \hp^{-1.1}$ and $\tn \appropto \hp^{-0.86}$ according to the fits (\ref{eq:fit-l}) \& (\ref{eq:fit-nl}), which are reasonably close to $\tau\propto \hp^{-1}$ in terms of the slope of scaling.


\subsection{Vortensity threshold for RWI}
\label{sec:zeta_th}


Whether the onset of planet-induced RWI can be directly associated with $\delta\zeta$ exceeding some threshold value $\zeta_\mathrm{th}$, as assumed in the derivation of the scaling (\ref{eq:estdzetadt}), is an important question. Existence of a well-defined $\zeta_\mathrm{th}$ would be extremely useful since then one could predict the emergence of vortices without running the time-consuming linear RWI analysis (it would still be needed if one were interested in e.g. the RWI growth rate); instead one would simply follow the evolution of the vortensity profile which can be easily done using the methods of \citetalias{Cimerman2021}. And it was shown in \citet{Ono2016} that for simple models of localized vortensity perturbations in discs (e.g. due to a Gaussian bump, or a sharp increase of $\Sigma$) some statements regarding the RWI triggering in terms of the amplitude and radial scale of $\delta \Sigma(R)$ features can indeed be made.

On the other hand, emergence of growing modes in equation (\ref{eq:mastereq}) is determined entirely by the behaviour of the potential $D$, given by equation (\ref{eq:D}). Even for simple models considered in \citet{Ono2016} $D$ has a very complicated radial structure, see Figs. 5 \& 9 in that work. In the planetary case $\zeta(R)$ is considerably more complex, see Fig. \ref{fig:rec_disc_q15}, which gets reflected in even more sophisticated radial profile of $D$. Moreover, RWI triggering depends not only on the amplitude but also on the width of vortensity features, which scales with $\lsh$ and changes as $\hp$ or $\Mp$ are varied \citep[also higher temperatures tend to promote RWI growth, see][]{Li2000}.

\begin{figure}
\centering
\includegraphics[width=0.49\textwidth]{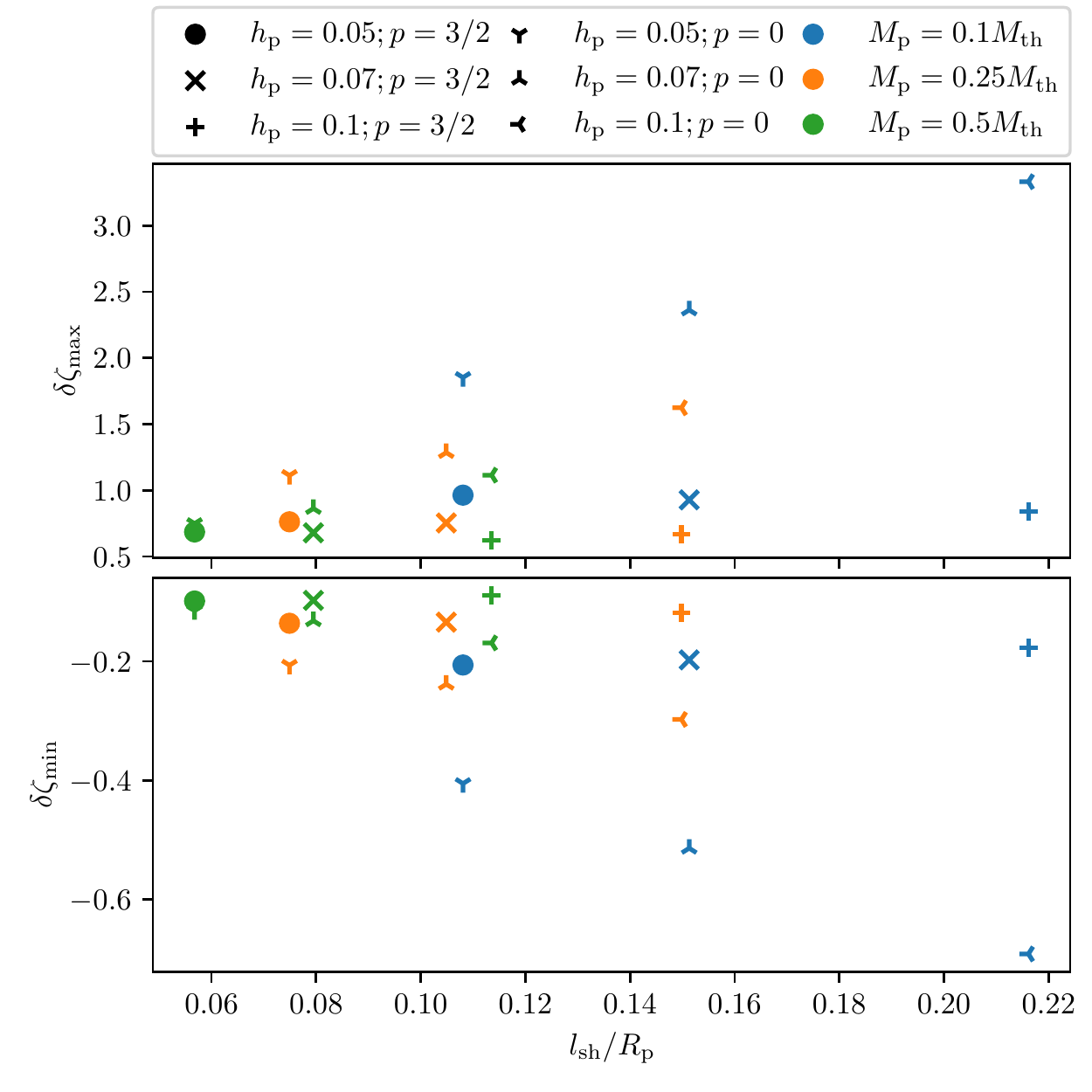}
\vspace*{-.3cm}
\caption{
Characteristic extremal values of vortensity perturbation (in units of $\Omega_\mathrm{p}/\Sigma_\mathrm{p}$) at the onset of linear RWI (at time $\tlrsa$), extracted for different disc-planet parameter sets (indicated with various colours and symbols in the legend). We show both the highest (top panel) and lowest (bottom panel) values of the vortensity perturbation at the edge of the gap where the instability develops first (which is determined by $p$ in this case). This calculation uses our semi-analytical framework.}
\label{fig:ext-zeta-lin}
\end{figure}

\begin{figure}
\centering
\includegraphics[width=0.49\textwidth]{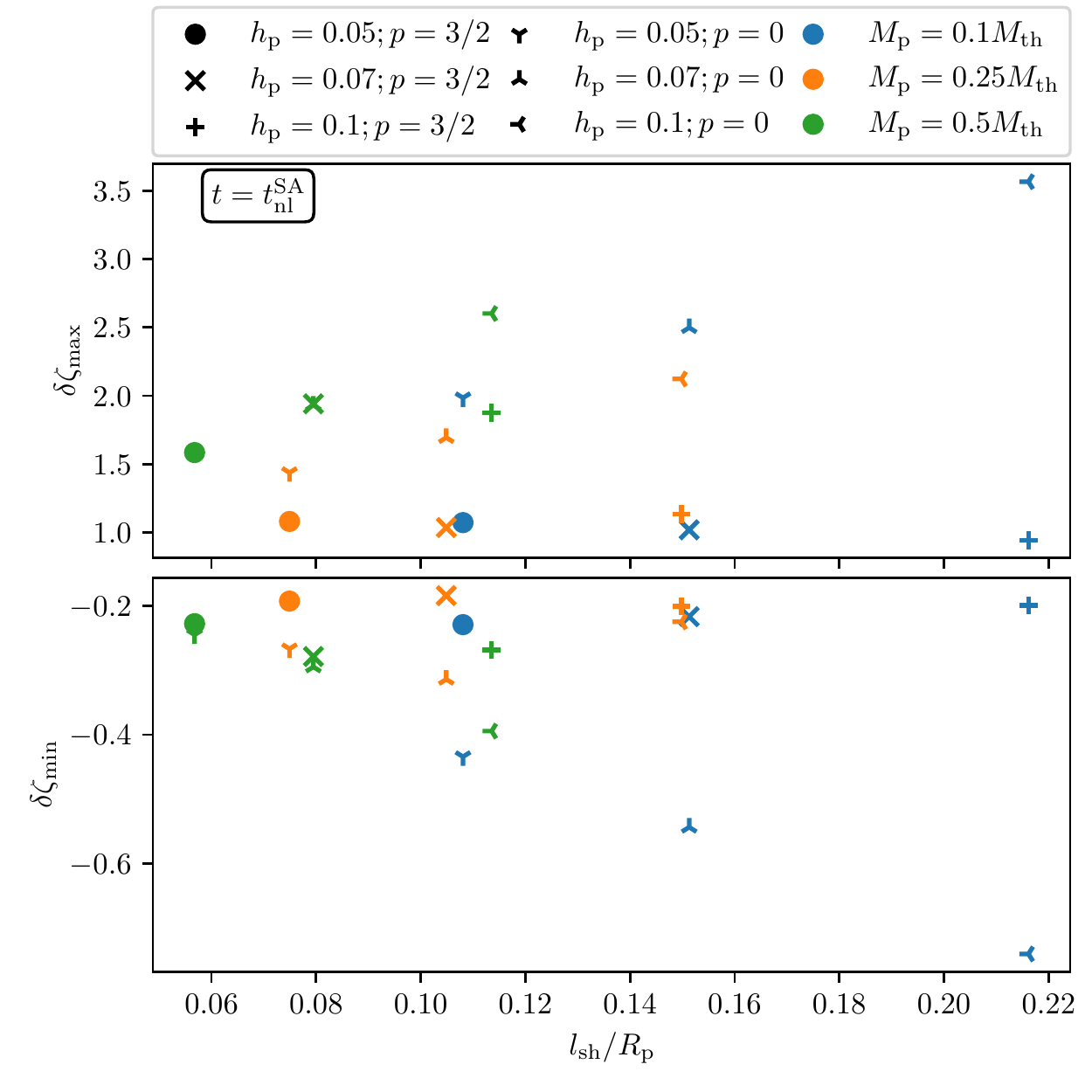}
\vspace*{-.3cm}
\caption{
Same as Fig. \ref{fig:ext-zeta-lin}, but now showing the extrema of vortensity perturbation at time $t_\mathrm{nl}^\mathrm{SA}$, when the RWI reaches its non-linear stage and fully developed vortices appear in simulations.}
\label{fig:ext-zeta-nonlin}
\end{figure}

Given these complications, to provide a direct test of our $\zeta_\mathrm{th}$ assumption, we measured the extremal values of $\delta\zeta$ (deviation from $\zeta_\mathrm{i}$) at time $\tlrsa$ (onset of RWI) in our semi-analytic\footnote{We verified that extrema of $\delta\zeta$ measured in simulations agree with these values typically to $\lesssim 10\%$, which should be obvious from Figs. \ref{fig:rec_disc_q15} \& \ref{fig:rec_disc_all}.} calculations for different disc and planet parameters. The results are shown as a function of $\lsh$ in Fig. \ref{fig:ext-zeta-lin}, where we display both the peak ($\delta\zeta_\mathrm{max}$, top) and trough ($\delta\zeta_\mathrm{min}$, bottom) values of $\delta\zeta$ on the side of the gap where the RWI sets in first (inner for $p=0$, outer for $p=3/2$). In Fig. \ref{fig:ext-zeta-nonlin} we plot the same information but at time $t_\mathrm{nl}^\mathrm{SA}$, when the RWI becomes non-linear and vortices should appear.

There are several observations that we can make based on these plots. First, extremal values of $\delta\zeta$ do depend on $\lsh$ --- the scale of vortensity features; at any given $\lsh$ there is a large scatter in $\delta\zeta$ (a factor of several) as different parameters are varied, but the envelope clearly increases with $\lsh$. Second, extremal values of $\delta\zeta$ follow well-defined branches when only one parameter of the model is varied. In particular, we show in Appendix \ref{sec:zeta_th_fits} that the data shown in Fig. \ref{fig:ext-zeta-lin} can be matched using a simple formula (\ref{eq:fit_zeta_ext_lin}) with three parameters depending on $p$. Thus, despite the complexity of the radial behaviour of $D$, the extrema of $\delta\zeta$ change in a systematic (and not chaotic) fashion. Third, as one decreases $\Mp/\Mth$ while keeping $\hp$ fixed (varying colour of symbols), the extremal $\vert\delta\zeta\vert$ increase, but much faster for $p=0$ than for $p=3/2$. Fourth, as one increases $\hp$ while keeping $\Mp/\Mth$ fixed, the extrema of $\vert\delta\zeta\vert$ {\it increase} for $p=0$, while slightly {\it decreasing} for $p=3/2$. Fifth, the extremal values of $\delta\zeta$ typically vary much less for $p=3/2$ (a factor of $\sim 2$) than for $p=0$ (which vary by up to $\sim 6$).

Given these trends, it appears that the stability of the gap edge to RWI is determined not only by the overall amplitude of the vortensity perturbation but also by other factors, e.g. the radial scale and overall shape of $\delta\zeta$. For that reason, it is also not surprising that the timescale fit parameters (\ref{eq:fit-l}) and (\ref{eq:fit-nl}) show some  deviations from the predictions of our simple estimate (\ref{eq:estdzetadt}). What is remarkable, is that these deviations are rather small and are almost independent of the surface density slope $p$, especially given the clearly $p$-dependent variation of the critical $\delta\zeta$ values (by a factor of several) and the diverse trends with $\hp$ and $\Mp$ in Figs. \ref{fig:ext-zeta-lin} \& \ref{fig:ext-zeta-nonlin}.

The larger amplitude of extremal $\delta\zeta$ for $p=0$ can be explained by the fact that in this case vortices first appear at the inner edge of the planetary gap. The rapid variation of $\vert\Omega-\Omega_\mathrm{p}\vert$ in the inner disc (see Fig. \ref{fig:lin_syn}) tends to broaden the $\delta \zeta(R)$ distribution, requiring higher values of $\delta \zeta$ for RWI to set in when $\lsh$ is large. But this also explains why the instability timescale stays roughly the same regardless of $p$: for large $\lsh$ the rate at which the fluid passes through the shock in the inner disc can be considerably higher than predicted by the local approximation (\ref{eq:linsyn}), see Fig. \ref{fig:lin_syn}. Our results imply that this effect almost fully compensates for the increased $\vert\zeta_\mathrm{th}\vert$ in $p=0$ discs, resulting in $\tlr$ and $\tn$ being essentially independent of $p$.

To summarize, we do not find that, in general, $\zeta_\mathrm{sh}$ is confined within a well-defined narrow range, as assumed in Section \ref{sec:theorytau}. This is approximately true for $p=3/2$, where we observe variations of $\zeta_\mathrm{th}$ by a factor of only $\sim 2$, but for $p=0$ critical $\zeta_\mathrm{sh}$ may vary much more. This is likely good news for interpreting observations, since we expect higher $p$ (i.e. radially decreasing $\Sigma$) to be more typical for PPDs. On the other hand, for any $p$ we do observe that the characteristic values of $\vert\zeta_\mathrm{th}\vert$ are of order unity, which is reassuring and may explain the insensitivity of both $\tlr$ and $\tn$ to $p$, and their rough agreement with the scaling (\ref{eq:estdzetadt}), which we found in Sections \ref{sec:time-lin} \& \ref{subsec:tauhat}. We leave detailed exploration of the trends observed in Figs. \ref{fig:ext-zeta-lin} \& \ref{fig:ext-zeta-nonlin} and \ref{fig:ext-zeta-lin-fit} to future work. 


\section{Discussion}
\label{sec:discuss}


Semi-analytic calculation of the timescales $\tlrsa$ and $t_\mathrm{nl}^\mathrm{SA}$ (see Section \ref{sec:RWI-time}) is a multi-step procedure: we first need to compute the evolution of the vortensity profile, then reconstruct $\Sigma,\Omega$ profiles (Section \ref{sec:SA_disc}), after which we perform the RWI stability analysis (Section \ref{sec:RWI_eqs}) and integrate growth rates to obtain $t_\mathrm{nl}^\mathrm{SA}$ (Section \ref{sec:RWI-time}). Despite this complexity, $t_\mathrm{nl}^\mathrm{SA}$ agrees remarkably well with the simulation-based $\thsim$, providing strong support for the robustness of our semi-analytical framework. We now discuss some ways in which one can capitalize on this agreement, and put our results in context with existing studies.


\subsection{Applications of our results}
\label{subsec:appl}


Our results (\ref{eq:fit-nl})-(\ref{eq:T_vrt}) for the planet-induced vortex development timescales can be used for interpreting observations of protoplanetary discs showing evidence of non-axisymmetric features (arcs, clumps, etc.), which may be interpreted as vortices. Without going into too many details here, we note that equation (\ref{eq:T_vrt}) can be used to constrain either the mass or the age of a planet responsible for the appearance of an observed vortex in a protoplanetary disc. We provide more details on this in Rafikov \& Cimerman (in prep.).

Our timescale estimates (\ref{eq:taufit})-(\ref{eq:fit-nl}) are also useful for planning future numerical studies of RWI and vortices in protoplanetary discs. They allow one to estimate the run time of the simulations producing vortices in advance, facilitating making informed decisions about allocation of computing resources. Our timescale fits (\ref{eq:taufit})-(\ref{eq:fit-nl}) can also be used for benchmarking numerical codes.

Our $\Sigma$, $\Omega$ reconstruction technique (see Section \ref{sec:SA_disc}) based on \citet{Lin2011} allows one to construct radial profile of the planet-induced gap given the knowledge of the radial vortensity profile. Once combined with our recipe for vortensity production (valid for $\Mp\lesssim \Mth$) it provides a fully self-contained semi-analytical method for predicting the radial profiles of shallow gaps at arbitrary moments of time, which we thoroughly test against simulations (see Section \ref{sec:comp_sa_sim} and Appendix \ref{sec:tests-SA}). This is a powerful tool, complementary to the existing methods for construction of profiles of deep gaps carved out by massive ($\Mp\gtrsim \Mth$) planets \citep{Crida2006,Kanagawa2017,Ginz2018,Duffell2020}, that can find multiple uses. 

In particular, radial profiles of $\Sigma$, which can be easily computed as a function of time using our method, could be used to study dust trapping at pressure bumps \citep[e.g. ][]{Stammler2019}. Especially at the outer gap edge, where our semi-analytical model matches simulations well, we expect modelling of dust accumulation to be quite accurate (in the inner disc multiple spirals may lead to some discrepancies with simulations). This provides an efficient and accurate way of studying e.g. the dependence of pebble isolation mass \citep{Paardekooper2004,Lamb2014} on disc properties such as $\hp$ and $p$ (in the inviscid limit).

Numerical studies of RWI development, dust accumulation, etc. near the low mass (sub-thermal) planets typically need to be run for thousands of orbits to properly capture the disc structure at late times. Our semi-analytical method provides a useful shortcut: one can use it to construct the approximate disc state (including the forming gap) around the planet at any moment of time after the planet has been introduced in the simulation. This state would provide the initial condition for running the simulation, avoiding the initial `burn-in' stage needed to reach this situation in the simulation.

We should also note that our semi-analytical method can naturally account for various processes that have been neglected in this study --- planet migration, history of its mass accumulation \citep{Hammer2017,Hallam2020}, evolution of the disc state on long timescales, etc. To enable this, one simply needs to generalize equation (\ref{eq:syn}) by allowing $\Omega_\mathrm{p}$ and $\Delta\zeta(R)$ to be functions of time set by the evolution of the physical characteristics of the problem at hand, e.g. planetary semi-major axis $\Rp$, mass $\Mp$, etc. This equation can then be easily integrated, obtaining a more general $\zeta(R,t)$ than the solution (\ref{eq:zeta_t}) used in this work\footnote{This implicitly assumes that the time-variation of $\Delta \zeta$ is slow compared to the dynamical time (which is typically satisfied, e.g. since our $t_\mathrm{lin}\gg P_\mathrm{p}$, see Figure \ref{fig:taulin}), such that the radial force balance, i.e. equation (\ref{eq:radHSE}), holds.}. Using this $\zeta(R,t)$ to reconstruct the evolving $\Sigma$, $\Omega$ profiles (see Section \ref{sec:SA_disc}) and running the RWI linear stability analysis (see Sections \ref{sec:RWI_eqs} \& \ref{sec:resRWI}) with these inputs, one can determine the onset of RWI in these more general situations.

For illustration, let us sketch how our heuristic determination of the vortex emergence timescale $\tau$ (see Section \ref{sec:theorytau}) would change if $\Mp$ were not constant but was actually growing in time due to accretion \citep{Hallam2020}. With time-dependent $\Mp(t)$ the condition (\ref{eq:inst_cond}) would become $\zeta_\mathrm{th}\approx \int^\tau_0\Ssh(t)\de t \propto \int^\tau_0\Mp^{2.6}(t)\de t$, see equation (\ref{eq:dzdt}), providing a new relation for determining $\tau$ instead of equation (\ref{eq:estdzetadt}). Because of the steep dependence on $\Mp(t)$ in this integral we expect that the determination of $\tau$ would be most sensitive to the details of the $\Mp(t)$ behaviour at late time, when the planet is close to reaching its final mass.


\subsection{Potential limitations of this work}
\label{sec:limits}


In this work, we made several simplifying assumptions to highlight the most important physical processes leading to vortex formation. In particular, we neglected the presence of dust in the disc and its back-reaction on the fluid motions and ignored the disc self-gravity \citep{Lin2012II}. Unlike \citet{Hammer2017} and \citet{Hallam2020}, we keep the planetary mass fixed in time since accounting for the planetary accretion history would necessarily introduce some ad hoc assumptions about the $\Mp(t)$ dependence. Similarly, we fix the semi-major axis of the planetary orbit not allowing it to migrate and set planetary eccentricity to zero. We also allow only one planet to be present in the disc. As mentioned in Section \ref{subsec:appl}, our semi-analytical approach outlined in Section \ref{sec:SA_disc} can be easily generalized to account for the history of planetary mass accretion and migration, multiple planets \citep{Carrido2022}, and so on. We restrict the disc to be strictly two-dimensional. However, in their study of linear RWI in a 3D disc, \citet{Lin2012I,Lin2012II} found that its growth rates can be accurately predicted from the 2D problem alone. Thus, introducing the third dimension is unlikely to change our main conclusions. 


\subsubsection{Non-barotropic thermodynamics}
\label{subsec:therm}


In line with our previous work
\citepalias{Cimerman2021}, all our calculations assume a globally isothermal EoS, which is a special case of a barotropic EoS ($P=P(\Sigma)$ only). In real discs there are various heating and cooling processes that would cause departures from the barotropic setup. Adoption of more sophisticated thermodynamic assumptions, even as simple as the introduction of $\beta$-cooling on a timescale $\beta\Omega_\mathrm{K}^{-1}$, would have several consequences.

First, the excitation of planetary density waves would change, see \citet{Miranda2020I} who explored the dependence of planetary torques on $\beta$. Second, (linear) thermal relaxation can dramatically modify damping of the spiral waves as shown in \citet{Miranda2020I,Miranda2020II}. This must affect radial distribution of the vortensity production around the planet and the resulting gap profiles. Third, baroclinic effects may start affecting vortensity evolution. Fourth, irreversible heating of the disc by planet-driven spiral shocks \citep[e.g.][]{Rafikov2016, Ziampras2020I} can modify the local temperature profile near the planet, further impacting density wave propagation and potentially driving additional baroclinic effects. Fifth, a specific form of the EoS affects the vortensity generation at the shock front, e.g. compare \citet{Lin2010} and \citetalias{Cimerman2021}. Finally, even for identical surface density perturbations (i.e. neglecting the aforementioned effects), the linear growth of the RWI is affected by thermal relaxation as shown by e.g. \citet{Les2015} and \citet{Huang2022_RWICooling}.

Thermal relaxation also affects the long-term survival of vortices \citep{FungOno2021,Rometsch2021}, but this issue is beyond the scope of our study, which focuses on vortex {\it generation}.


\subsubsection{The effect of shear viscosity}
\label{sec:visc}


In line with \citetalias{Cimerman2021}, we assumed the disc to be inviscid. This assumption is supported by many observations suggesting that viscosity is likely low in most PPDs \citep{Pinte2016,Rafikov2017,Flaherty2020}. Nevertheless, it is still important to assess the consequences of disc viscosity being non-zero.

\citet{Miranda2020I} have shown that, unless the disc is very viscous (with effective $\alpha\gtrsim 0.01$), viscous stresses do not affect density wave dissipation, which is determined by the combination of non-linear and radiative damping. However, another effect of viscosity is to diffusively smooth out any features in the radial vortensity distribution. In particular, it introduces the diffusive term in the vortensity evolution equation, which will tend to smear out the sharp peaks and troughs of $\zeta$ produced in the vicinity of the planet, see Figs.  \ref{fig:rec_disc_q15} \& \ref{fig:rec_disc_all}. As a result, viscosity would slow down the growth of $\zeta$ and could delay the onset of RWI as has been shown in e.g. \citet{Hallam2020}. Thus, our inviscid calculations provide us with the {\it lower limit} for the time to reach instability, and $\tau$ will be longer in sufficiently viscous discs.

High viscosity may suppress the RWI entirely \citep{Hammer2017}, by preventing $\zeta$ from reaching the threshold value $\zeta_\mathrm{th}$ necessary for instability to set in. We can provide a simple heuristic estimate of when this might happen. Without invoking the explicit form of the viscous term in the vortensity evolution equation, we can model it as a diffusion process\footnote{It should be remembered that the real stress in the disc, for example caused by the MRI, may respond to planetary torques differently from a simple shear viscosity model \citep[e.g.][]{Zhu2013}.} with a characteristic diffusion coefficient equal to the kinematic viscosity $\nu$ (similar to the momentum equation), i.e. $\partial_t\zeta\vert_{\nu}\sim \nu\nabla^2\zeta$. Since $\zeta$ varies on scales $\sim\lsh$, this term should be of order $\nu\zeta/\lsh^2$. Viscous diffusion would stop the planet-driven growth of vortensity at some equilibrium value $\zeta_\mathrm{eq}$ at which $\partial_t\zeta\vert_{\nu}$ would match $\Ssh$ given by equation (\ref{eq:dzdt}). In other words, 
\begin{align}
\Omega_\mathrm{p} \hp \left( \frac{\Mp}{\Mth} \right)^{2.6}\propto \nu \frac{\zeta_\mathrm{eq}}{\lsh^2},
\end{align}	 
which, using equation (\ref{eq:lsh}) and $\nu=\alpha\hp^2 \Omega_\mathrm{p}\Rp^2$, can be written as
\begin{align}
\zeta_\mathrm{eq}\propto \alpha^{-1}\hp\left( \frac{\Mp}{\Mth} \right)^{1.8}.
\label{eq:zeta_eq}
\end{align}
RWI would be suppressed altogether and vortices would not emerge if $\zeta_\mathrm{eq}<\zeta_\mathrm{th}$. In particular, equation (\ref{eq:zeta_eq}) implies that for this to be the case planetary mass must satisfy
\begin{align}
\frac{\Mp}{\Mth} < B \left(\frac{\alpha}{\hp}\right)^{5/9},   
\label{eq:visc_lim}    
\end{align}
where $B$ is a (dimensionless) constant which can be calibrated using simulations.
This relation is necessarily approximate, with true power law exponents possibly deviating somewhat from our predictions, which again can be checked using simulations. We note that \citet{McNally2019} found a different relation between $\Mp$ and $\alpha$ (with the opposite sign of the exponent), which separates evolution with and without vortices. The reason for this difference is that (unlike us) they also considered planet migration, which can lead to non-trivial feedback effects in the presence of viscosity, see \citet{Rafikov2002II}. We also cannot directly compare our relation (\ref{eq:visc_lim}) with the results of \citet{Hallam2020}, since planet mass was varied in their study. In light of our results, formation of vortices in viscous discs clearly warrants further investigation.


\subsection{Comparison with previous works}
\label{subsec:compprev}


A number of past studies explored the emergence of vortices at the edges of planetary gaps, and some of their approaches share similarities with our work. For example, in their study of high mass planets (their lowest mass is $\Mp = 0.8 \Mth$), \citet{dvB2007} investigated linear stability to RWI of a simulated disc perturbed by a planet in a time-dependent manner, like we do when determining $\tlrsim$ and $\tlsim$. They found an increase in the growth rate of unstable modes with time (see their Fig. 5) as the planet opens a deeper and deeper gap, similar to what we show in our Fig. \ref{fig:RWI_growthrates_q15}. Adopting a constant $\Sigma_\mathrm{i}$ (i.e. $p=0$) disc, they found higher growth rates at the inner gap edge, in line with our findings. While they speculated that this is caused by the proximity to domain boundaries, our analysis shows that this is a real effect due to the accelerated steepening of the inner spiral arm for $p=0$ \citepalias{Cimerman2021}.

\citet{Lin2010} carried out a similar analysis, also focusing on higher mass planets ($\Mp \geq \Mth$). Their work was the first to introduce $\Sigma, \Omega$ reconstruction from vortensity $\zeta$, and they found good performance of this technique using $\zeta$ profiles from simulations. We extend this method even further by employing a semi-analytic prescription for $\zeta$ evolution from \citetalias{Cimerman2021}, see Section \ref{sssec:comp_sa_sim}, and confirm its accuracy for a range of relevant parameters, see Section \ref{sec:comp_sa_sim} and Appendix \ref{sec:tests-SA}. 

A number of studies \citep[e.g.][]{Li2009,Yu2010,McNally2019} numerically studied emergence of vortices in viscous discs with migrating planets. Because of these additional physical ingredients, which introduce new effects \citep[e.g. gas redistribution associated with migration,][]{Rafikov2002} we cannot provide a direct comparison of our results with their findings. This also precludes us from comparing our calculations of the vortex emergence timescale $\tau$ with the works of \citet{Hammer2017} and \citet{Hallam2020}, since they considered time-varying planet mass and viscous discs (although we discuss ways in which the impact of these additional physical ingredients might be understood in Sections \ref{subsec:appl} and \ref{sec:visc}, respectively).

Finally, we note the similarity of planet-driven vortex production in PPDs with the origin of vortices found by \citet{Coleman2022} in their inviscid 2D simulations of accretion disc boundary layers. The only key difference is that in their case density waves are driven not by planets but by the acoustic instability in the boundary layer \citep{Belyaev2012,Belyaev2012b,Belyaev2013}, while all subsequent processes --- production of vortensity at the shock fronts, triggering of RWI, mergers of the resultant vortices --- are essentially identical.


\section{Summary}
\label{sec:summ}


We studied the stability of gaps carved by (sub-thermal mass, fixed in time) planets to RWI in globally isothermal, inviscid, 2D protoplanetary discs. Our primary goals were to study the disc evolution leading to instability and to determine the timescale on which RWI sets in and vortices form at the edges of planet-driven gaps. We used a two-stage approach  to reach these goals. First, we developed a closed-form semi-analytical approach for computing the radial profile of a gap induced by a sub-thermal mass planet at any moment of time given a set of disc and planetary parameters (Section \ref{sec:SA_disc}). This was achieved by coupling the semi-analytical calculation of the vortensity production at the planetary shock \citepalias{Cimerman2021} with the $\Sigma,\Omega$-reconstruction technique \citep{Lin2010}. Second, we carried out a linear RWI stability analysis (Section \ref{sec:RWI_eqs}) on these gap profiles to determine the onset of instability. Along the way, our results have been verified at all levels against direct hydro simulations. Below we briefly summarize our main findings.

\begin{itemize}

\item We showed  (Section \ref{sec:comp_sa_sim} and Appendix \ref{sec:tests-SA}) that our semi-analytical method reliably reproduces gap profiles when compared to expensive 2D simulations for a mass range $\Mp = (0.1 - 0.5) \Mth$ (with somewhat reduced agreement in the inner disc caused by the emergence and dissipation of the secondary spiral arm). This allowed us to study the RWI stability of gap edges in a fast and efficient manner. 

\item We calculated the timescales for the onset of linear RWI and for the development of its non-linear phase. We provided useful fitting formulae for these timescales (Sections \ref{sec:time-lin},\ref{subsec:tauhat}) valid in the $\Mp\lesssim \Mth$ regime and proposed an approximate, heuristic explanation of these scalings (Section \ref{sec:theorytau}). 

\item Consistent with previous studies, we found that hotter discs (keeping $\Mp/\Mth$ constant)
and higher mass planets accelerate the development of RWI.

\item In agreement with \citetalias{Cimerman2021}, we found  that the disc surface density slope $p$ controls, which side of the gap develops RWI first: inner gap edge for constant-$\Sigma$ discs ($p=0$), but outer edge for discs with radially decreasing $\Sigma$ ($p=3/2$). 

\item We found that at the point when RWI starts developing, the planet-induced vortensity deviation $\delta\zeta$ shows considerable variation as disc and planetary parameters are changed (Section \ref{sec:zeta_th}). While we observe certain trends in the behaviour of this characteristic $\delta\zeta$, its high sensitivity to the value of the surface density slope $p$ precludes us from drawing simple and universal conclusions. 

\item We have shown how our semi-analytical approach can be extended beyond our simple setup, for example, to account for planetary accretion and migration (Section \ref{subsec:appl}), and assessed the impact of non-zero disc viscosity on our results (Section \ref{sec:visc}).

\item Our findings can be used for interpreting observations of PPDs with vortex-like structures that may be caused by planets, as well as for setting up and testing simulations of disc-planet interaction. 

\end{itemize}

In Rafikov \& Cimerman (in prep.) we apply the results of this study to constrain masses and ages of putative planets in protoplanetary discs with vortex-like structures.


\section*{Acknowledgements}

\textit{Software:} NumPy \citep{2020NumPy-Array}, SciPy \citep{2020SciPy-NMeth}, IPython \citep{IPython}, Matplotlib \citep{Matplotlib}, Athena++ \citep{Athenapp2020}.
We thank the referee Takayuki Muto for a constructive report which helped us in clarifying several points made in this work.
N.P.C. would like to thank Robin Croft for helpful discussions regarding the relaxation solver and all developers of Athena++ for making their code publicly available. N.P.C. is funded by an Isaac Newton Studentship and a Science and Technology Facilities Council (STFC) studentship.
R.R.R. acknowledges financial support through the NASA grant 15-XRP15-2-0139, Ambrose Monell Foundation, and STFC grant ST/T00049X/1. A large part of the long term simulations were performed on the HPC cluster FAWCETT at DAMTP, University of Cambridge. Part of this work was performed using resources provided by the Cambridge Service for Data Driven Discovery (CSD3) operated by the University of Cambridge Research Computing Service (\texttt{www.csd3.cam.ac.uk}), provided by Dell EMC and Intel using Tier-2 funding from the Engineering and Physical Sciences Research Council (capital grant EP/P020259/1), and DiRAC funding from the Science and Technology Facilities Council (\texttt{www.dirac.ac.uk}).

\section*{Data Availability}
The data underlying this article will be shared on reasonable request to the corresponding author.




\bibliographystyle{mnras}
\bibliography{example} 




\appendix

\section{Search for RWI modes}
\label{app:find}
We discretize the differential operators in equation (\ref{eq:RWI}) using $N_R$ cells in radius, which allows us to write this equation as
\begin{align}
	W(R,\omega,m) \Psi_m = 0,
	\label{eq:RWI-eig}
\end{align}
where $W$ is a complex, tri-diagonal square matrix of size $N_R \times N_R$. Existence of non-trivial solutions $\Psi_m$ thus requires $\det W = 0$ or, in other words, that $W$ has at least one eigenvalue $\lambda_i = 0$. The problem is then reduced to finding the values of $\omega$ giving $\det W = 0$. For most parameter sets, we solve this problem on a subdomain of the same logarithmic radial grid that is used in the hydrodynamical models, ranging from $R_\mathrm{min}=0.3 R_\mathrm{p}$ to $R_\mathrm{max}=3.25 R_\mathrm{p}$. For some parameter sets (i.e. small $\Mp$ and large $\hp$), this radial domain is increased to ensure that boundaries are far enough from the gap edges.

In order to allow for wave propagation across boundaries, we use outgoing wave boundary conditions under the WKB approximation \citep{Ono2016}. While low-$m$ modes ($m$ $\lesssim 6 $) are very localized and insensitive to boundary conditions, modes of higher $m$ ($\gtrsim 6 $) are more global, such that BCs become important \citep[e.g.][]{Lin2010}.

We use an eigenvalue search strategy similar to \citet{Li2000}, \citet{dvB2007} and \citet{Ono2016}, which we have implemented using \texttt{Python3}. Knowing the radial location of the vortensity minimum $R_0$, we use the result of previous studies that trapped, unstable modes are corotation modes, i.e. we start our search around $\omega_\mathrm{R} \simeq m \Omega (R_0)$ (and $\vert \Delta \omega \vert^2 \ll \Omega^2$). We thus restrict our search to a region around corotation, typically $0.9 < \omega_\mathrm{R} / (m \Omega (R_0)) < 1.1$ and begin by looking for modes with low growth
rates $0.01 < 2 \pi \gamma_m < 0.4$. Poles and roots of $\det W$ are found via contour integration in the complex $\omega_\mathrm{R}-\gamma$ plane as described in \citet{dvB2007}. When a contour contains a root, we use Muller's method \citep{Muller1956} for locating it more precisely. Once a mode is found, $\Psi_m$ is obtained by solving the eigenvalue problem (\ref{eq:RWI-eig}) using the SciPy routine \texttt{scipy.sparse.linalg.eigs} for sparse matrices (since $W$ is tri-diagonal). Especially for modes with low $\gamma_m$, the solver sometimes finds spurious unphysical modes, that do not represent trapped modes. Thus, our method requires some human interaction, but could be extended to deal with such spurious solutions. Having found an unstable mode for the disc structure at a time $t$, we can use the complex frequency as an initial guess in Muller's method for the next search at time $t + \Delta t$, as the disc evolves slowly \citep{Li2000}. This saves the more expensive contour integration time.

We have verified our method by confirming that we find the same modes and growth rates as \citet{Ono2016} for their fiducial `Gaussian Bump' setup. As another test, we have run several simulations in which we removed planet after the disc reached an unstable state. Re-initializing the disc with the azimuthally averaged $\langle \Sigma(R) \rangle_\phi$, $\langle \Omega(R) \rangle_\phi$ from the last frame with the planet, we monitored the development of RWI with a well-defined (constant in time) growth rate. In parallel, we also carried out linear RWI analysis on the same background disc state. We found very good agreement (at the level of a few percent) of both the growth rates and the radial structure of the mode measured in simulations with linear theory. This experiment not only verifies our linear RWI analysis but also provides a very useful test of the code, in particular, of the orbital advection module, now implemented in Athena++, which has not been used by \citet{Ono2018}.

\begin{figure*}
\centering
\includegraphics[width=\textwidth]{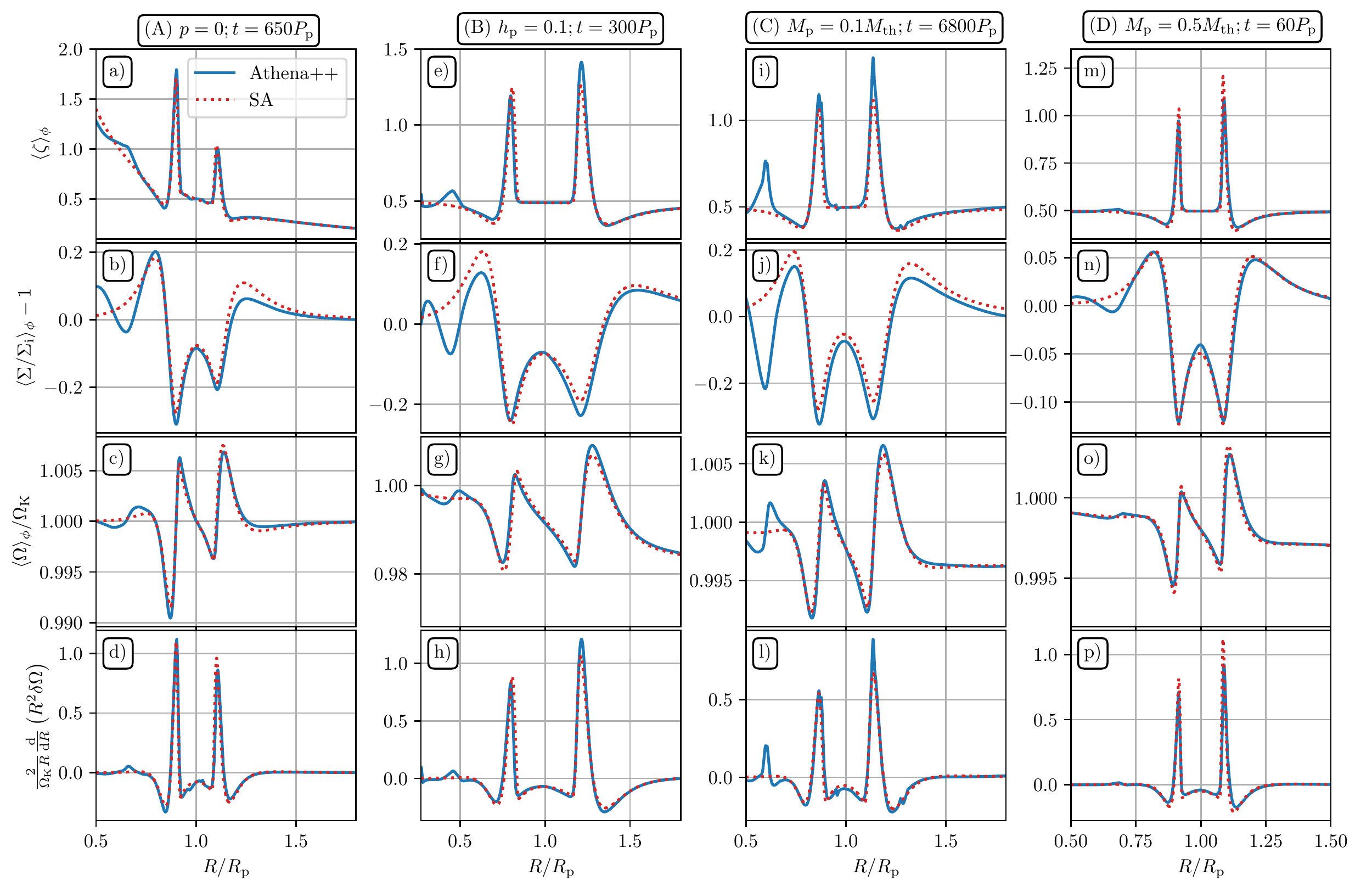}
\vspace*{-0.7cm}
\caption{
Same as Fig. \ref{fig:rec_disc_q15}a,c,d,e, but for different sets of parameters and times (what is different compared to the fiducial set $\Mp = 0.25\Mth; \hp =0.05; p=3/2$ is indicated above each column), comparing near-gap $\zeta$, $\Sigma$, and $\Omega$ profiles from simulations and the semi-analytical method. See Appendix \ref{sec:tests-SA} for details.
}
\label{fig:rec_disc_all}
\end{figure*}


\section{$\Sigma,\Omega$ reconstruction}
\label{app:vortHSE}


To solve their version of equation (\ref{eq:ODEsigma}), \citet{Lin2010} adopted a shooting method.
However, we found that it fails if the radial region of integration is too large, as only exponentially growing or decaying solutions are found.

For this reason, we developed a more robust iterative method for solving this non-linear differential equation: we used a discretized version of equation (\ref{eq:ODEsigma}),
\begin{align}
	\frac{\Sigma_{i+1} - 2 \Sigma_{i} + \Sigma_{i-1}}{w^2} = F_i(\Sigma,\zeta,R),
\end{align}
where $w \equiv R_{i+1} - R_i$ is a constant grid spacing and subscript $i$ indicates evaluation at $R_i$ and
\begin{align}
F_i(\Sigma,\zeta,R) = &\frac{2\zeta \Sigma_{i}^2}{\cs^2}\left(\Omega_\mathrm{K}^2 + \frac{\cs^2 \left(\Sigma_{i+1}-\Sigma_{i-1}\right)}{2 w R \Sigma_{i}}\right)^{1/2}\\ 
	&-\frac{\Omega_\mathrm{K}^2}{\cs^2} \Sigma_{i} +\frac{\left(\Sigma_{i+1}-\Sigma_{i-1}\right){}^2}{4 w^2 \Sigma_{i}}
	- \frac{3 \left(\Sigma_{i+1}-\Sigma_{i-1}\right)}{2 w R}
	 \nonumber .
\end{align}
This can be rearranged to give an iterative scheme for $\Sigma_i$:
\begin{align}
	\Sigma_i^{(n+1)} = (1-C) \Sigma_i^{(n)} + \frac{C}{2} \left(-w^2F_i + \Sigma_{i+1} + \Sigma_{i-1} \right)^{(n)},
\end{align}
where the superscript indicates the number of iteration. We typically chose $C = 2/3$. If required, adopting a multi-grid method and making $C$ a function of iteration number for successive over-relaxation (SOR) could accelerate convergence.


\section{Further tests of the semi-analytical reconstruction}
\label{sec:tests-SA}


Here we describe further tests of our semi-analytical gap reconstruction technique (Section \ref{sec:SA_disc}), for disc and planetary parameters different from the fiducial setup.


\subsection{Variation of the surface density slope $p$}
\label{sec:vardslope}


First, we test reconstruction in a constant surface density disc ($p=0$). In column A of Fig. \ref{fig:rec_disc_all},
we show the disc structure at $t = 650 \Pp$ for our standard $\Mp = 0.25 \Mth$ and $\hp = 0.05$ but a constant $\Sigma_\mathrm{i}$, i.e. $p = 0$. As pointed out in \citetalias{Cimerman2021}, this disc model has a non-zero background vortensity gradient (compared to the fiducial $p=3/2$ disc), such that $\zeta_\mathrm{i}$ is no longer radially constant. The most important effect of the density slope $p$ is that it controls the part of the disc (inner or outer), in which the vortensity jump at the shock $\Delta \zeta$ is greater, as the non-linearity of the spiral wake is modified \citepalias[e.g.][]{Rafikov2002,Cimerman2021}. Accordingly, the inner gap edge experiences faster vortensity evolution (and becomes unstable earlier) than the outer one in a $p=0$ disc, opposite to the $p=3/2$ case covered in Section \ref{sssec:comp_sa_sim}. This is what we see in Fig. \ref{fig:rec_disc_all}a.

As in the fiducial case, vortensity and rotation profiles show good agreement between the semi-analytical method and simulations. Panel (b) reveals that the semi-analytical method matches $\Sigma$ at the inner gap edge well, but over-predicts it at the outer gap edge. This is due to a consistently stronger perturbation in $\zeta^\mathrm{SA}$ in this region. Both methods show a stronger asymmetry in the $\Sigma$ peaks at the inner and outer gap edges as compared to the fiducial case, with the inner peak dominating. The radial width of the gap is similar to the fiducial case, as expected, since $\lsh$ is unchanged.

Overall, the level of accuracy of our semi-analytical reconstruction for $p=0$ disc is similar to that in Fig. \ref{fig:rec_disc_q15}. We note, however, that for the lowest planet mass case, $\Mp = 0.1 \Mth$, with $\hp = 0.1$ and $p=0$, we find poorer agreement between simulations and semi-analytical method in the inner disc. We attribute this to the fact that in our semi-analytical model, we neglect the advection term in the conservation equation for vortensity, as we found it to be negligible in \citetalias{Cimerman2021}. However, for this parameter set, the shocking distance $\lsh$ is large, such that gap opening in the inner disc occurs in a region with initially large vortensity gradient. As a result, the vortensity advection term ($v_R \partial_R \zeta$) might become important as disc fluid gets radially redistributed. This argument is supported by the fact that we do not see such disagreement in the $p=3/2$ disc with radially constant $\zeta_\mathrm{i}$ (with other parameters kept the same).


\subsection{Variation of the disc scale-height $\hp$}
\label{sec:varhp}


We next change the disc scale-height at the planet location $\hp$ (and $\cs$), considering hotter discs. In column B of Fig.  \ref{fig:rec_disc_all} we compare our gap reconstruction with simulations at $t = 300 \Pp$ for fiducial values of the planet mass and surface density slope ($\Mp = 0.25 \Mth$ $p=1.5$) but an increased $\hp = 0.1$. Again, we find good agreement between our semi-analytical method and simulations.

In line with our findings in \citetalias{Cimerman2021} (e.g. Fig. 15 therein), the radial scale (width) of vortensity and density perturbation induced by the spiral shocks increases roughly linearly with $\hp$ since $\lsh\propto \hp$ for fixed $\Mp/\Mth$. While the vortensity jump at the shock $\Delta \zeta$ remains almost unchanged, the rate of vortensity production scales with the relative (synodic) orbital period of gas parcels with respect to the shock, see equation (\ref{eq:syn}). This means that vortensity perturbations increase more rapidly in hotter discs (as $\lsh$ is larger), leading to higher surface density perturbations at the same time after introduction of the planet: for $\hp =0.1$, $\delta \Sigma/\Sigma_0 \sim 0.1$ at the gap edges is reached after about half the time it takes in the $\hp=0.05$ disc.


\subsection{Variation of the planet mass $\Mp/\Mth$}
\label{sec:varmp}


The amplitude of the vortensity jump at the shock strongly depends on the normalized planet mass \citepalias[close to $\Delta \zeta \propto \left( \Mp/\Mth \right)^3$,][]{Cimerman2021}. Also, according to the equation (\ref{eq:lsh}), the width of the gap (which is $\propto\lsh$) decreases as $\Mp/\Mth$ increases. Both these scalings conspire to produce steeper gradients in surface density and pressure, and thus increased shear, as $\Mp/\Mth$ increases.

In columns C and D of Fig. \ref{fig:rec_disc_all} we show results for the fiducial disc parameters with a lower mass planet $\Mp = 0.1 \Mth$ at $t = 6800 \Pp$ and a higher mass planet $\Mp = 0.5 \Mth$ at $t = 60 \Pp$. Comparing the two illustrates the aforementioned trends. Note that the significance of the secondary gap in the inner disc (which causes vortensity to deviate from our reconstructed $\zeta$-profile, see a peak in $\zeta$ derived from simulations at $R=0.6$ in Fig. \ref{fig:rec_disc_all}i) relative to the primary gap is decreasing as $\Mp/\Mth$ increases. This is due to the fact that the density wave produced by a higher mass planet is more non-linear and loses its angular momentum to the background flow more efficiently. This leads to a less efficient formation of the secondary
spiral arm \citepalias[e.g.][]{Cimerman2021}.

Note that regardless of the disc and planetary parameter choices we always find the second term in the equation (\ref{eq:zeta_dev}), shown in panels (d), (h), (l), (p), to greatly exceed the first term in that equation, $\delta\Sigma/\Sigma_\mathrm{i}$ (panels (b), (f), (j), (n)). This emphasizes once again the importance of properly accounting for the small variations of $\Omega$ when computing $\zeta(R)$.


\section{RWI analysis for a $p=0$ disc}
\label{sec:RWI-another}


\begin{figure}
\centering
\includegraphics[width=0.49\textwidth]{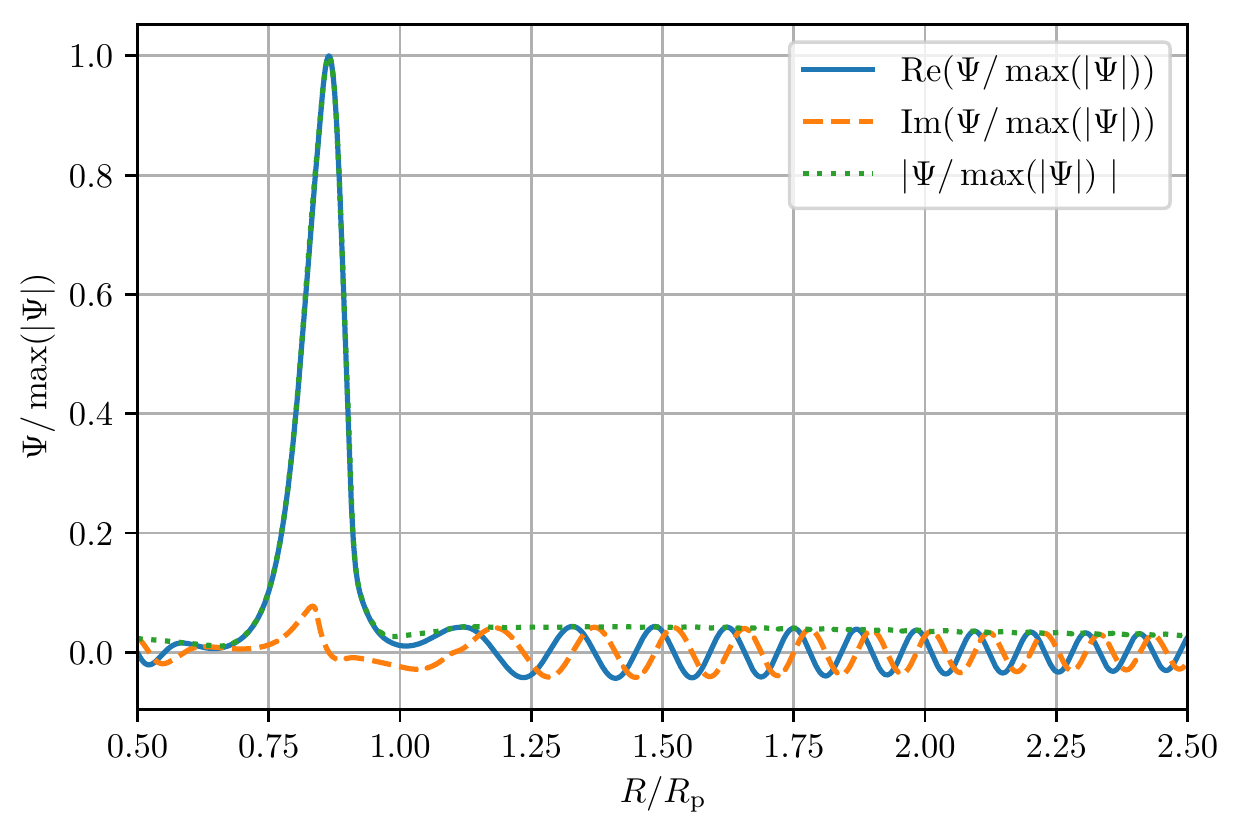}
\vspace*{-.6cm}
\caption{Same as the top panel Fig. \ref{fig:m3_sim_mapdP} but for an inner gap edge mode in the $p=0$ disc at $t = 650 \Pp$. The frequency of this mode is $\omega = 3.874+0.01534\I$.}
\label{fig:m3_I_sim_mapdP}
\end{figure}

\begin{figure}
\centering
\includegraphics[width=0.49\textwidth]{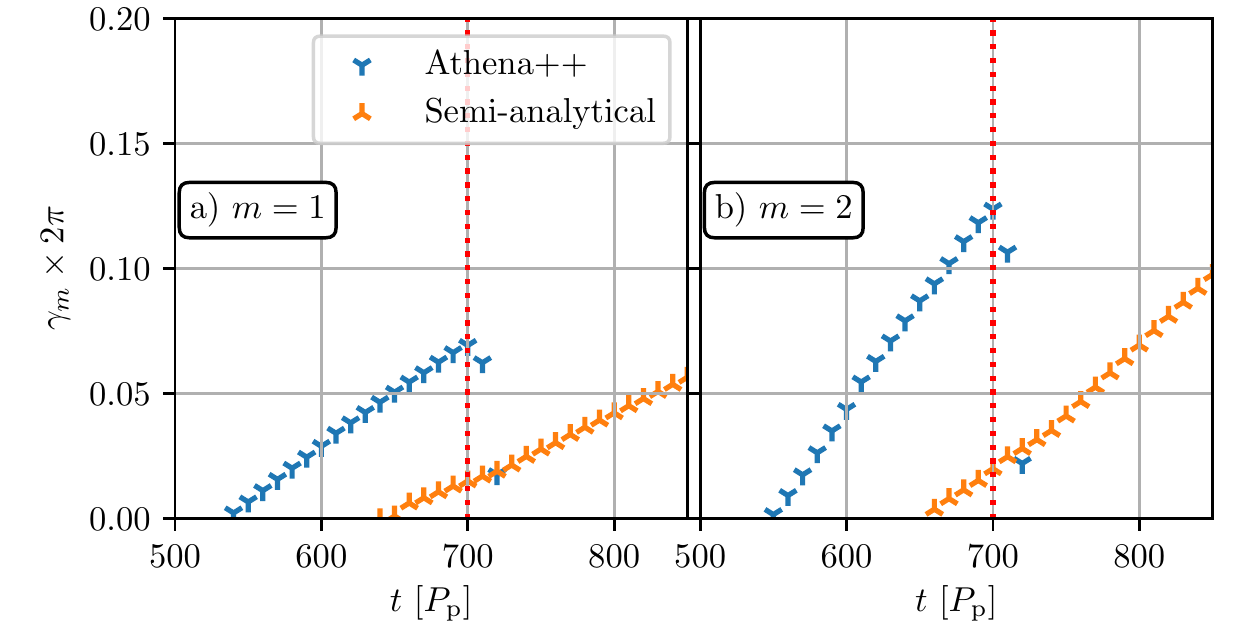}
\vspace*{-0.45cm}
\caption{Growth rates $\gamma$ of unstable modes, associated with the inner gap edge, found at different times for the constant surface density disc $(p=0$) and $\Mp/\Mth = 0.25$. Compare with Figure \ref{fig:RWI_growthrates_q15}.
}
\label{fig:RWI_growthrates_q0}
\end{figure}

In a constant surface density $(p=0)$ disc, the inner gap edge becomes unstable first as the vortensity evolves faster there, see column A of Fig. \ref{fig:rec_disc_all}. Figure \ref{fig:m3_I_sim_mapdP} illustrates this by showing the radial profile of the dominant $m=2$ eigenmode in $p=0$ disc at $t=650\Pp$. One can see that it peaks at $R\approx 0.85 \Rp$, i.e. at the inner gap edge, meaning that this side of the gap turns RWI-unstable first.

Due to the presence of the secondary spiral shock in the inner disc, not captured by our semi-analytical model, we might then also expect greater deviations of the growth rates from simulations in $p=0$ disc. Fig. \ref{fig:RWI_growthrates_q0} confirms this expectation by showing growth rates from our linear stability analysis for $m=1$ and $m=2$ modes. While the general behaviour of the growth rates (their overall increase with time) is similar to Fig. \ref{fig:RWI_growthrates_q15}, the time offset between the rates derived using the semi-analytical reconstruction (yellow) and azimuthally-averaged simulation data is larger in the $p=0$ disc, around $100\Pp$.


\section{Vortensity threshold for RWI onset}
\label{sec:zeta_th_fits}

\begin{figure}
\centering
\includegraphics[width=0.49\textwidth]{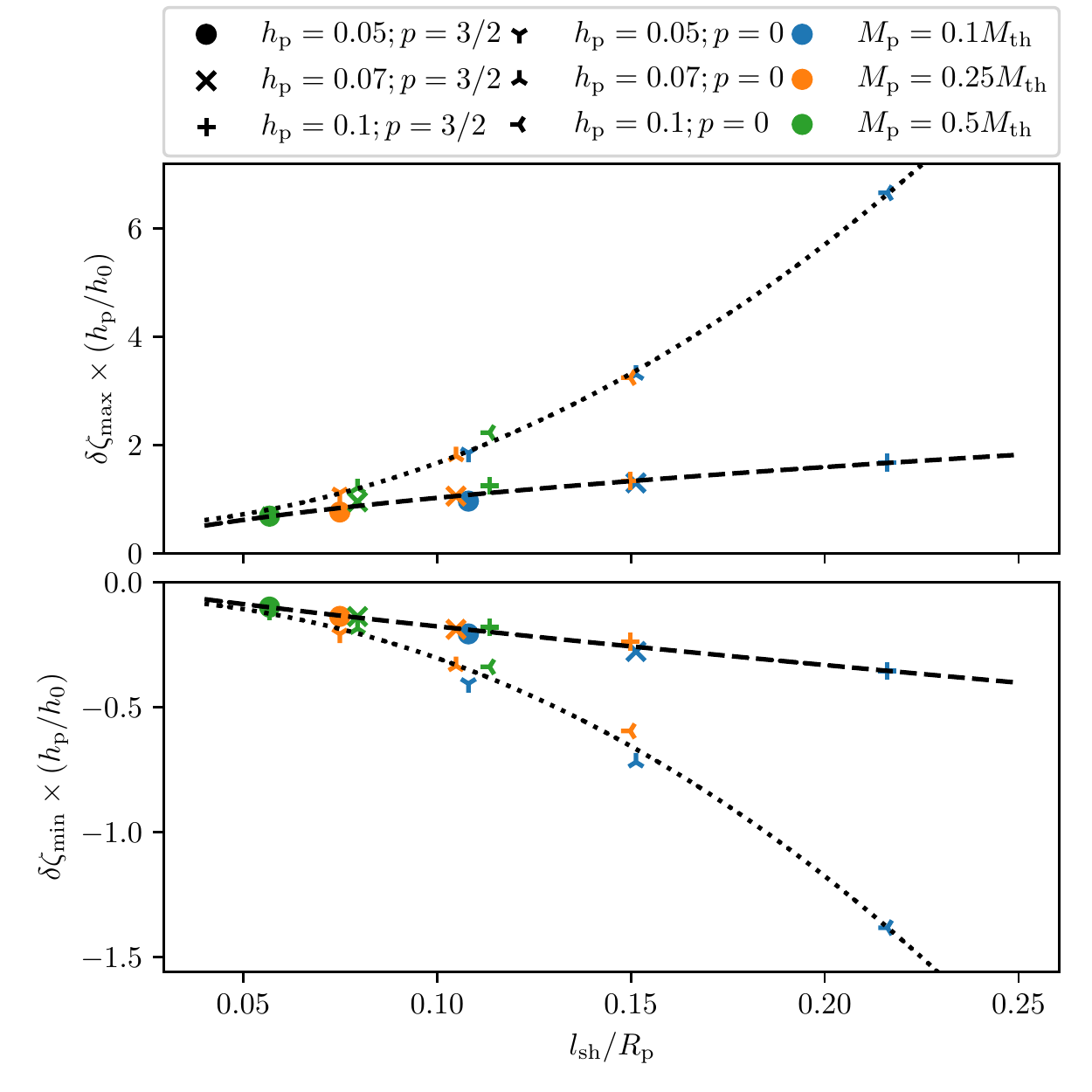}
\vspace*{-.6cm}
\caption{Data for $\delta \zeta_\mathrm{th}^\mathrm{lin}$ from Fig. \ref{fig:ext-zeta-lin} rescaled via multiplication by $\hp/h_0$ ($h_0=0.05$). One can see the points collapsing on two well-defined branches for different values of $p$. A fit (\ref{eq:fit_zeta_ext_lin}) to these data with parameters from Table \ref{tab:fit_zeta_ext} is shown by black dotted (for $p=0$) and dashed (for $p=3/2$) curves.}
\label{fig:ext-zeta-lin-fit}
\end{figure}

\begin{table}
\centering
\caption{Fitting parameters for the threshold vortensity perturbation at $t = \tlrsa$ (see equation \ref{eq:fit_zeta_ext_lin} and Fig. \ref{fig:ext-zeta-lin-fit}).
}
\label{tab:fit_zeta_ext}
  \begin{tabular}{lcccc}
	\hline
    \hline
    & $p$ & $a_1$ & $a_2$ & $a_3$ \\
	\hline
    \hline
	$\delta\zeta_\mathrm{min}$ & 3/2 & 0.035 & -1.3 & 0.8 \\
    & 0 & -0.05 & -36 & 2.2 \\
	\hline
	$\delta\zeta_\mathrm{max}$ & 3/2 & -0.43 & 4.4 & 0.48 \\
    & 0 & 0.43 & 153 & 2.1\\
    \hline
	\hline
  \end{tabular}
\end{table}

We found that multiplication by $\hp$ causes $\delta \zeta_\mathrm{th}^\mathrm{lin}$ data from Fig. \ref{fig:ext-zeta-lin} to fall onto two very well-defined branches as a function of $\lsh$, distinguished by the corresponding $p$. This is shown in Fig. \ref{fig:ext-zeta-lin-fit}. There we also show a fit of the form
\begin{align}
    \delta \zeta_\mathrm{th}^\mathrm{lin} = \left[a_1 + a_2\, \left(\frac{\lsh}{\Rp}\right)^{a_3} \right]
    \left(\frac{\hp}{h_0}\right)^{-1},
    \label{eq:fit_zeta_ext_lin}
\end{align}
where $h_0 = 0.05$ and $(a_1,a_2,a_3)$ are the fit parameters obtained via a least-square fit, done separately for the two values of $p$ and for the minima and maxima of vortensity at the inner (outer) gap edge for $p=0$ ($3/2$), giving us four sets of these coefficients. These parameter sets are listed in Table \ref{tab:fit_zeta_ext}.

The correlations found in Fig. \ref{fig:ext-zeta-lin-fit} clearly merit further investigation, especially for other $p$, to determine the dependence of $(a_1,a_2,a_3)$ on $p$. For now, we refrain from attaching any significance to these trends beyond the statements made in Section \ref{sec:zeta_th}.


\bsp	
\label{lastpage}
\end{document}